\documentclass[apj, numberedappendix]{emulateapj}
\usepackage{verbatim}
\usepackage{natbib}

\def\del#1{{}}

\newcommand{\mathbfit}[1]{\textbf{\textit{#1}}}

\newcommand{\rmn}{\mathrm}
\newcommand{\dd}{\mathrm{d}}

\newcommand{\bl}{\begin{large}}
\newcommand{\el}{\end{large}}

\newcommand{\M}{{\mathcal M}}

\newcommand{\eps}{\varepsilon}
\newcommand{\vecbf}{\mathbfit}

\newcommand{\apgt} {\ {\raise-.5ex\hbox{$\buildrel>\over\sim$}}\ }
\newcommand{\aplt} {\ {\raise-.5ex\hbox{$\buildrel<\over\sim$}}\ } 

\shorttitle{NGC 1265 and the Perseus Accretion Shock}
\shortauthors{Pfrommer \& Jones}

\begin{document}
\title{Radio Galaxy NGC 1265 unveils the Accretion Shock onto
  the Perseus Galaxy Cluster}

%
\author{
C.~Pfrommer\altaffilmark{1} and
T.~W.~Jones\altaffilmark{2}}

\altaffiltext{1}{Heidelberg Institute for Theoretical Studies, %
  Schloss-Wolfsbrunnenweg 35, D-69118 Heidelberg, Germany; %
  Canadian Institute for Theoretical Astrophysics, %
  University of Toronto, Toronto, ON, M5S~3H8, Canada; christoph.pfrommer@h-its.org}
\altaffiltext{2}{School of Physics and Astronomy, University of Minnesota, %
                  Minneapolis, MN 55455, USA; twj@msi.umn.edu}


\begin{abstract}
  We present a consistent three-dimensional model for the head-tail radio galaxy
  NGC 1265 that explains the complex radio morphology and spectrum by a past
  passage of the galaxy and radio bubble through a shock wave. Using analytical
  solutions to the full Riemann problem and hydrodynamical simulations, we study
  how this passage transformed the plasma bubble into a toroidal vortex
  ring. Adiabatic compression of the aged electron population causes it to be
  energized and to emit low surface brightness and steep-spectrum radio
  emission. The large infall velocity of NGC 1265---which is barely
  gravitationally bound to the Perseus cluster at its current position---and
  the low Faraday rotation measure values and variance of the jet strongly argue
  that this transformation was due to the accretion shock onto Perseus situated
  roughly at $R_{200}$. Estimating the volume change of the radio bubble enables
  inferring a shock Mach number of $\M\simeq 4.2_{-1.2}^{+0.8}$, a density jump
  of $3.4_{-0.4}^{+0.2}$, a temperature jump of $6.3_{-2.7}^{+2.5}$, and a
  pressure jump of $21.5\pm10.5$ while allowing for uncertainties in the
  equation of state of the radio plasma and volume of the torus. Extrapolating
  X-ray profiles, we obtain upper limits on the gas temperature and density in
  the infalling warm-hot intergalactic medium of $kT\lesssim0.4$~keV and
  $n\lesssim 5\times 10^{-5}\,\rmn{cm}^{-3}$. The orientation of the
  ellipsoidally shaped radio torus in combination with the direction of the
  galaxy's head and tail in the plane of the sky is impossible to reconcile with
  projection effects. Instead, this argues for post-shock shear flows that have
  been caused by curvature in the shock surface with a characteristic radius of
  850~kpc. The energy density of the shear flow corresponds to a
  turbulent-to-thermal energy density of 14\%---consistent with cosmological
  simulations. The shock-injected vorticity might be important in generating and
  amplifying magnetic fields in galaxy clusters.  We suggest that future
  polarized radio observations by, e.g., LOFAR of head-tail galaxies can be
  complementary probes of accretion shocks onto galaxy clusters and are unique in
  determining their detailed flow properties.
\end{abstract}


 

\section{Introduction}

Head-tail radio galaxies show spectacular asymmetric radio morphologies and
occur in clusters of galaxies.  The favored interpretation of these head-tail
sources is radio jets ejected by an active core of a galaxy. The jets are bent
at some angle toward one direction giving rise to a ``head'' structure and fan
out at larger distances in a characteristic tail that extends over many tens to
hundreds of kpc. If the flow impacting these galaxies is supersonic, ram
pressure of the intracluster medium (ICM) causes the jets to bend
\citep{1979Natur.279..770B}; if the flow is trans-sonic with a Mach number
$\M\sim1$, the thermal pressure gradient of the interstellar medium of these
galaxies due to their motion through the ICM causes the bending
\citep{1979ApJ...234..818J}.  In this model, the supersonic inflow is
decelerated and heated by a bow shock in front of the galaxy which also
generates a turbulent wake that re-accelerates the relativistic particle
population in the tail and illuminates the large-scale tail structure of these
sources. The Perseus cluster is an outstanding example, as it hosts many of
these head-tail sources: most prominently NGC 1265 and IC 310
\citep{1968MNRAS.138....1R, 1975A&A....38..381M}. High-resolution observations
of the head structure of NGC 1265 reveal two radio jets emerging from the galaxy
which are at an angle of $90\degr$ to the tail where the two jets are first
parallel to each other and then merge \citep{1986ApJ...301..841O}.
\citet{1979A&A....76..109G} found an extension of this radio tail to the
north-east. Surprisingly, at lower surface brightness,
\citet{1998A&A...331..901S} find an additional large scale structure that arches
toward the east around the steep spectrum tail of NGC 1265 and forms a closed
ring with a very steep, but constant spectral index across. Using rotation
measure (RM) synthesis, \citet{2005A&A...441..931D} find a very weak diffuse
polarized structure with a Faraday depth of approximately $50~\rmn{rad\,m}^{-2}$
at the angular position coincident with the steep spectrum tail as well as the
ring structure of NGC 1265. Follow-up mosaic observations of the area around the
Perseus cluster show that most if not all of this emission is of Galactic origin
\citep{2011A&A...526A...9B}.

\citet{1998A&A...331..901S} discuss a total of four models to explain this
puzzling observation. They immediately discard two models---one model of chance
superposition of several independent head-tail galaxies due to the lack of
strong radio sources in this field (Model 1) as well as another model that
hypothesizes reacceleration of mildly relativistic electrons in the turbulent
wake of a galaxy due to contrived projection probabilities and implausible
energetics (Model 2).  The previously favored pair of models discussed in
\citet{1998A&A...331..901S}---apparently simple---do in fact appear to have
considerable amount of fine-tuning and have severe physical problems associated
with them as we will point out here.  Model 3: in this model, they postulate a
helical cluster wind that has to be aligned with the line-of-sight (LOS) to
produce the observed ring structure in projection: such a configuration is
highly unlikely, and this model has problems in explaining the constancy of the
spectrum and the surface brightness along the radio ring. To balance the
synchrotron and inverse Compton cooling of an electron population, this model
would need to postulate a fine-tuned acceleration process that however must not
fan out the well-confined radio emission along the arc. Model 4: in this model,
the ``radio tail'' would outline the ballistic orbit of NGC 1265. To sustain such
a helical orbit of NGC 1265 over $360\degr$, this model would require an
undetected dark object of mass $M\gtrsim M_\rmn{NGC~1265}\simeq
3\times10^{12}M_\odot$ orbiting the galaxy.\footnote{Since the velocity
  dispersion of NGC 1265 is unknown, we use the Faber-Jackson relation in
  combination with the $M_\rmn{bh}$--$\sigma$ relation \citep{2002ApJ...574..740T}
  to obtain the stellar mass of NGC 1265 of $M_\ast\simeq
  2.5\times10^{11}\,M_\odot$. We estimate the total mass by using the
  universal baryon fraction and a stellar-to-baryon mass yield of 0.5 so that
  the resulting halos mass should be considered a lower limit to the true one.}
The change of the direction in the brighter part of the tail remains
unexplainable and there is also the problem in explaining the constancy of the
spectrum and the surface brightness along the radio ring.  Given these
difficulties, it is attractive to consider alternative possibilities that may
explore the interaction between a radio galaxy with the outskirts of the Perseus
ICM. This has been foreshadowed in a remark by \citet{2005A&A...441..931D} who
speculate whether the large scale polarized structure that arches around the
steep spectrum tail of NGC 1265 is indeed the remains of an earlier phase of
feedback. Our work will demonstrate that this picture is not only the simplest
consistent explanation for the radio morphology and spectrum, but we also use it
to indirectly infer the presence of a cluster shock wave and measure its
properties.

Previously, the morphology of a giant radio galaxy has already been used to
indirectly detect a large scale shock at an intersecting filament of galaxies
\citep{2001ApJ...549L..39E} by using the radio galaxy as a giant cluster weather
station \citep{1998Sci...280..400B}. Gravitationally driven, supersonic flows of
intergalactic gas follow these filaments toward clusters of galaxies that
represent the knots of the cosmic web \citep{1996Natur.380..603B}.  The flows
will inevitably collide and form large-scale shock waves
\citep{1998ApJ...502..518Q, 2000ApJ...542..608M, 2003ApJ...593..599R,
  2006MNRAS.367..113P}. Of great interest is the subclass of accretion shocks
that are thought to heat the baryons of the warm-hot intergalactic medium (IGM)
when they are accreted onto a galaxy cluster. Formation shocks have also been
proposed as possible generation sites of intergalactic magnetic fields
\citep{1997ApJ...480..481K, 1998A&A...335...19R, 2008Sci...320..909R}. Prior to
this work, only discrete merger shock waves have been detected in the X-rays
\citep[e.g.,][]{2002ApJ...567L..27M} and there was no characterization possible of
the detailed flow properties in the post-shock regime as to test whether shear
flows---the necessary condition for generating magnetic fields---are present.

The structure of the paper is as follows. In Section~\ref{sec:idea}, we present
the basic picture of our model in a nutshell while we derive the detailed
three-dimensional (3D) geometry of NGC 1265 within our model in
Section~\ref{sec:geometry}. In Section~\ref{sec:shock}, we work out the
properties of the accretion shock onto the Perseus cluster including those of
the post-shock flow and present the implications for the infalling warm-hot
IGM.  In Section~\ref{sec:model}, we carefully examine the
underlying physics as well as the hydrodynamic stability of our model and
discuss our findings in Section~\ref{sec:conclusions}.  Throughout this work, we
use a Hubble constant of $H_{0} = 70\,\rmn{km~s}^{-1}\rmn{Mpc}^{-1}$. For the
currently favored $\Lambda$CDM cosmology with the present day density of total
matter, $\Omega_m=0.28$, and the cosmological constant, $\Omega_\Lambda=0.72$,
we obtain an angular diameter distance to Perseus ($z=0.0179$) of
$D_\rmn{ang}=75\,\rmn{Mpc}$; at this distance, $1\arcmin$ corresponds to
$21.8\,\rmn{kpc}$. X-ray data estimates a virial radius and mass for
Perseus\footnote{We define the virial mass $M_\Delta$ and the virial radius
  $R_\Delta$ as the mass and radius of a sphere enclosing a mean density that is
  $\Delta=200$ times the critical density of the Universe.}  of
$R_{200}=1.9\,\rmn{Mpc}$ and $M_{200}=7.7\times10^{14}M_\odot$
\citep{2002ApJ...567..716R}.


\section{The idea in a nutshell}
\label{sec:idea}

\begin{figure*}
\epsscale{1.15}
\plotone{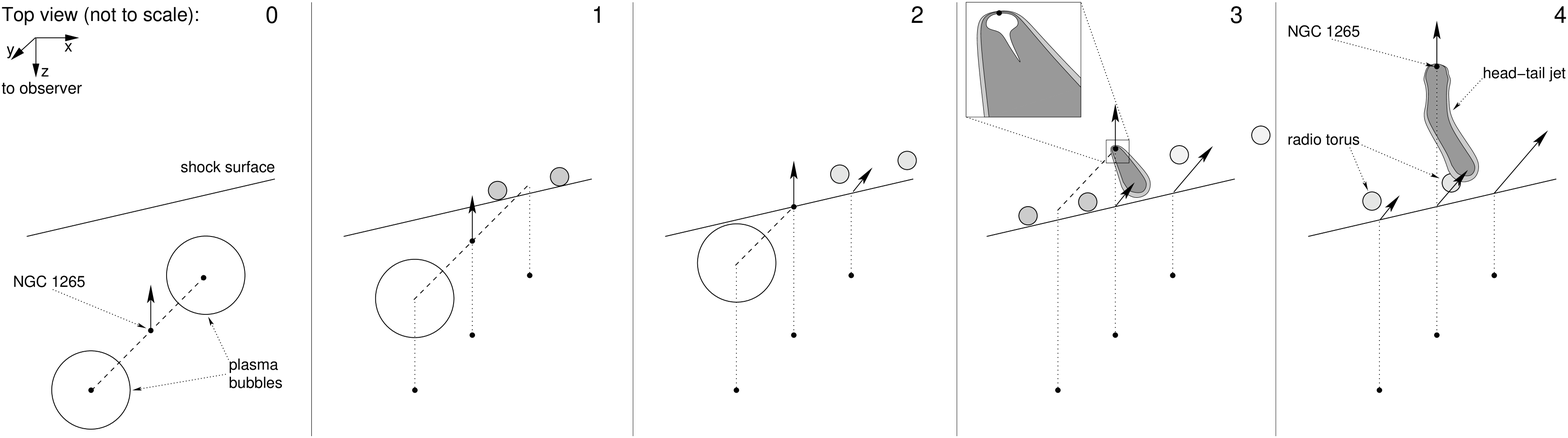}
\caption{Schematic snapshots of the time evolution of our model of NGC 1265
    (cut plane of the top view with the two axes representing the LOS and a
    suitably chosen angular direction, not to scale). Frame 0 shows the
    initial configuration of NGC 1265 together with two detached plasma bubbles
    from previous outbursts as they are moving upward toward the shock
    surface. In frame 1, the first bubble on the right passes through the
    shock and transforms into a torus (vortex ring) which re-energizes an aged electron
    population by means of adiabatic compression so that it emits observable
    radio synchrotron radiation. As shown in frame 2, its orbit is subject
    to the coherent shock deflection upon oblique shock crossing. In this frame,
    NGC 1265 also crosses the shock but remains on its ballistic orbit. However
    in our model, the shock passage triggers a new outflow that is now seen as
    head-tail radio emission. The jets are being bent by the ram pressure wind
    which causes them to merge into a single radio-emitting tail at low radio
    frequencies (shown in the schematic inlay in frame 3). Also shown in
    this frame is the transformation of the second plasma bubble on the left
    into a torus. Frame 4 shows today's configuration: the head-tail jet
    of NGC 1265 is shaped by post-shock flows owing to the oblique shock
    crossing, the second torus on the left is stretched by post-shock shear
    flows (Figure~\ref{fig3}) and emits diffuse soft radio emission, and the
    electron population in the first torus on the right (not drawn) has
    sufficiently cooled such that its radio emission is too weak to be detected
    by current radio telescopes.}
  \label{fig1}
\end{figure*}

Before we present the idea of our model, we point out the main morphological and
spectral properties of the giant radio galaxy NGC 1265 that every model would
have to explain: (1) The synchrotron surface brightness, $S_\nu$, and the
spectral index, $\alpha$, between 49 and 92 cm along the tail of NGC 1265
(starting at the galaxy's head) show a characteristic behaviour (as shown in
Figure~2 of \citealt{1998A&A...331..901S}). In the first part of the tail, both
quantities decline moderately in a way that is consistent with synchrotron and
inverse Compton cooling of a relativistic electron population that got
accelerated at the base or the inner regions of the jet. (2) At the point of the
tail where the twist changes in projection from left- to right-handed, both
quantities experience a sudden drop---while $S_\nu$ changes by a factor of 10,
$\alpha$ declines from $-1.1$ to $-2.1$. (3) Finally, along the remaining curved
arc, $S_\nu$ and $\alpha$ stay approximately constant on a total arc length of
$l_\rmn{arc} = 2\pi R \xi \sqrt{1 + k^2}\gtrsim 700 \,\rmn{kpc}$, where
$R\simeq150\,\rmn{kpc}$ is the radius of the arc, $\xi\simeq 3/4$ the projected
arc length in units of $2\pi$ radians, and $k=h/(2\pi R)\geq0$, where $h$ is the
height of the 3D helix. This property is in particular puzzling, as there is no
visible synchrotron cooling or fanning out of the dilute part of the tail
visible.

These findings taken together suggest the presence of two separate populations
of relativistic electrons giving rise to the bright and the dim part of the
tail, respectively, where the latter must have experienced a coherent
energetization event\footnote{We discard the alternative that there is a
  continuous acceleration process that exactly balances synchrotron and inverse
  Compton cooling due to the large degree of fine-tuning needed and because it
  is difficult to avoid a fanning out of the synchrotron emission region along
  the tail if the acceleration process is, e.g., of turbulent nature.}
over a length scale of $2 R\simeq 300\,\rmn{kpc}$ and on a timescale that is
shorter than the cooling time of the radio emitting electrons of
$\tau_\rmn{sync,\,ic} \lesssim 2.9\times 10^8\,\rmn{yr}$. The presence of one
radio tail that connects the two electron populations in projection points to a
causally connected origin of the synchrotron radiating structure. The most
natural explanation that combines these observational requirements are the
reminders of two distinct epochs of active galactic nucleus outbursts where the
most recent one is still visible as a head-tail radio jet and the older one
experienced a recent coherent energetization event. In particular, we propose
that such an energizing event could be provided by the passage of a detached
radio plasma bubble from a previous outburst through a shock wave.  This passage
transforms the plasma bubble into a torus (vortex ring) and adiabatically
compresses and energizes the aged electron population to emit low surface
brightness and steep-spectrum radio emission (as we detail below). If the shock
crossing is oblique, then the radio torus and the tail end of the recent
outburst experience the same degree of shock deflection since both structures
can be regarded as passive tracers of the post-shock velocity field and hence
provide a natural explanation for the apparent connectivity of both structures.
With this idea in mind, we sketch a schematic of the time evolution in our model
in Figure~\ref{fig1}.

\begin{figure*}
\begin{center}
\begin{minipage}{1.33\columnwidth}
  \includegraphics[width=\textwidth]{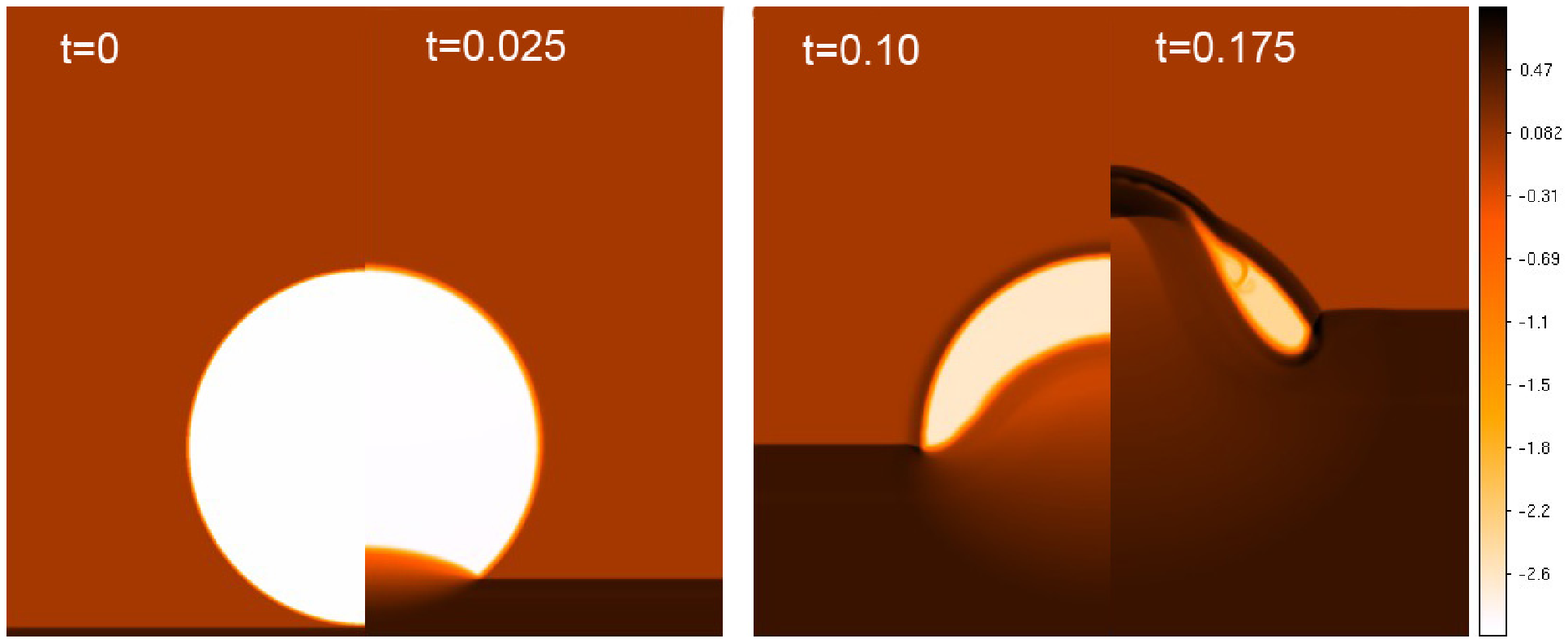}
\end{minipage}
\begin{minipage}{0.69\columnwidth}
  \vspace{0.5em}
  \includegraphics[width=\textwidth]{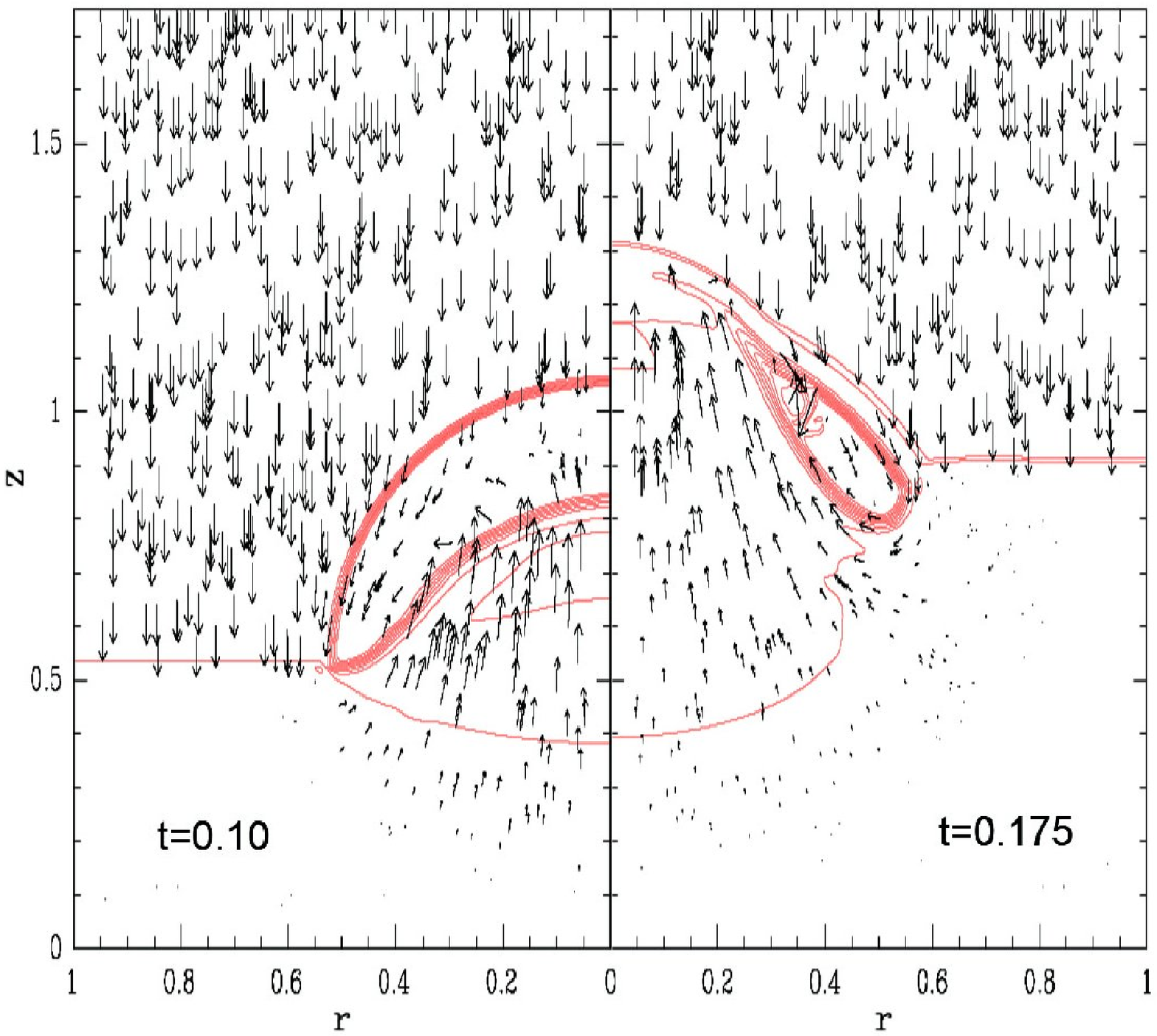}
\end{minipage}
\end{center}
\caption{Left two panels: images from a 2D axisymmetric simulation showing
  the evolution of a spherical bubble overrun by a Mach 5 shock that transforms
  into a torus (each half-panel shows a selected snap shot). Base 10 log of gas
  density is shown. The initial bubble/ICM density contrast $\delta = 10^{-3}$,
  the bubble diameter is unit length, the pre-shock ICM density is unity, as is
  the pre-shock sound speed.  Times correspond to initial shock impact on the
  bubble ($t=0$), just before the internal bubble shock has crossed the bubble
  (not visible with this color scale, $t=0.025$), the ICM shock has crossed a
  bubble radius (externally, $t=0.1$) and just after the bubble contact
  discontinuity has crossed the bubble ($t=0.175$).   Right panel: the
  velocity field in the rest frame of the ICM post-shock flow overlaid by gas
  density contours. Note that the stabilizing vortex flow around the torus has
  already established at the time when the torus has formed.}
\label{fig:images} 
\end{figure*}

The reason for the transformation of a plasma bubble into a toroidal vortex ring
can be easiest seen in the rest frame of the shock, where the ram pressure of
the pre-shock gas balances the thermal pressure in the post-shock regime. The
bubble is filled with hot (relativistic) and more dilute plasma compared to the
surrounding ICM.  Once the dilute radio plasma of the bubble comes into contact
with the shock surface, the ram pressure is reduced at this point of contact due
to the smaller density inside the bubble \citep{2002MNRAS.331.1011E}. The shock
and the post-shock gas expand into the bubble and propagate with a faster
velocity compared to the incident shock in the ICM. Owing to symmetry, the
ambient gas penetrates the line through the center of the bubble first and has a
smaller velocity for larger impact parameters.  This difference in propagation
velocities implies a shear flow and eventually causes a vortex flow around the
newly formed torus which stabilizes it as it moves now with the post-shock
velocity field.  To study the timescale on which this transformation happens, we
solve the one-dimensional (1D) Riemann problem of a shock passage through a
bubble exactly in Appendix~\ref{sec:Riemann}. These calculations are
complemented by a suite of two-dimensional (2D) axisymmetric simulations using a
code that employs an upwind, total variation diminishing scheme to solve the
hydrodynamical equations of motion \citep{1998ApJ...509..244R}. To map out
parameter space, we varied the Mach number and initial bubble/ICM density
contrast (see Figure~\ref{fig:images} for one realization). As a result, we find
that an initially spheroidal bubble will then evolve into a torus on a timescale
$\tau_\rmn{form}$ that is determined by the crossing time of the original
bubble--ICM contact discontinuity (CD) through the bubble.  For typical numbers,
$\tau_\rmn{form}\simeq 1.4\times 10^8\,\rmn{yr}$ (Equation~(\ref{eq:tau_form})).
As we will see in Section~\ref{sec:model}, this is much faster than any
transverse shear on the scale of the bubble or beyond can act to distort it.

  The morphology of the radio torus is consistent with MHD simulations of
this effect that assume strong shocks \citep[cf. Figs.~9 and 10
in][]{2002MNRAS.331.1011E}.  The relativistic plasma within the bubble
experiences an adiabatic compression by a factor\footnote{In this work we will
  encounter three different compression factors/density contrasts that should
  not be confused: the compression factor $C=V_\rmn{bubble}/V_\rmn{torus}$ by
  which the volume of the bubble changes upon shock passage, the compression
  factor $C_s=\rho_2/\rho_1$ between the post- and pre-shock ICM density at the
  shock front, and the density contrast $\delta=n_\rmn{torus}/n_\rmn{icm}$
  between the density of the radio plasma and the ICM.}  $C$ that increases the
Lorentz factor of the relativistic electrons as $\Gamma\propto C^{1/3}$ and the
rms magnetic field as $B\propto C^{2/3}$. Hence the radio cutoff of a
cooled electron population increases as $\nu_\rmn{max}\propto B \Gamma^2\propto
C^{4/3}$ and illuminates a previously unobservable radio plasma
\citep{2001A&A...366...26E}. Synchrotron and inverse Compton aging develops a
steep spectrum with spectral index $\alpha\sim2$ and a low surface brightness
$S_\nu\propto\nu^{-\alpha}$.

The galaxy itself remains on a ballistic orbit that is unaffected by the shock
passage that likely triggers a new outflow that is now seen as a head-tail radio
emission with the jets being bent by the ram pressure wind.  The associated
compression wave propagating through the interstellar medium could have
triggered a bar-like instability in the inner accretion disk that enabled
efficient angular momentum transport outward and accretion onto the
super-massive black hole (SMBH) which was responsible for launching the jet.

This model explains the observed spectral steepening of the head-tail radio
galaxy of $\Delta \alpha \simeq 0.5$ due to radiative cooling along the bright
part of the tail. In particular, our model naturally accounts for the
observed sudden steepening of the spectral index and dropping synchrotron
brightness at the transition (in projection) of the radio tail of the current
outburst to the shock-illuminated radio torus and match the observed constancy
of spectrum and surface brightness across the torus (see Figure~\ref{fig3} and
\citealt{1998A&A...331..901S}).

The shock compression acts mostly perpendicular to the filaments during the
formation of the torus. Consequently, the field component parallel to the bubble
surface is preferentially amplified and dominates eventually the magnetic energy
density \citep{2002MNRAS.331.1011E}. (Provided the initial magnetic field
component parallel to the shock was not too weak to be amplified above the
perpendicular components.)  Hence, our model predicts radial synchrotron
polarization vectors relative to the center of the radio torus with a
polarization degree of $\sim 5\%$ for our derived geometry of a viewing angle of
$\chi\simeq 23^\circ$ (see Section~\ref{sec:geometry}). This is consistent with
upper limits on the polarization of $10\%$--$25\%$
\citep{1998A&A...331..901S}. Future low-frequency radio observations with, e.g.,
LOFAR should be able to detect this signal.


\section{Observables and Geometry}
\label{sec:geometry}
 \begin{table*}[hbt!]
\begin{center}
\caption{Definitions}
\begin{tabular}{ll}
  \hline\hline
  $\vecbf{e}_r$,  $~\vecbf{e}_r^2=1$ & LOS vector pointing to observer\\
  $\vecbf{n}_s$,  $\,\vecbf{n}_s^2=1$ & Shock normal, pointing outward, away from convexly curved surface\\
  $\vecbf{v}$, $~~\vecbf{v}^2 = v^2$  & 3D proper motion of NGC 1265 relative to Perseus, pointing inward\\
  $\chi  =\arccos(\phantom{-}\vecbf{e}_r\cdot\vecbf{n}_s)$& ``Shock orientation,'' angle between LOS and shock normal\\
  $\theta=\arccos(-\vecbf{e}_r\,\cdot\frac{\vecbf{v}}{v})$& ``Inclination of the galaxy's orbit,''
  angle between negative LOS and galaxy velocity\\
  $\phi  =\arccos(-\vecbf{n}_s\cdot\frac{\vecbf{v}}{v})$& ``Shock obliquity,''
  angle between negative shock normal and galaxy velocity\\
  \hline
  Subscripts $t, r$ & Transverse (on the plane of the sky), radial (along LOS)\\
  Subscripts $\perp, \parallel$ & Perpendicular, parallel to shock normal\\
  Subscripts 1, 2 & Pre-shock, post-shock regime; refers to quantities of the gas\\
  \hline\hline
\end{tabular}
\label{tab:def}
\end{center}
\end{table*}

In this slightly technical section, we derive the specific geometry that we will
use later on to demonstrate the physical plausibility of our proposed model---if
the reader is only interested in the physical ideas and results, this section
might be easily skipped without loss of the underlying logic of this work.

In general, the location and orientation of the shock surface with respect to
the galaxy's infall velocity vector and the projection geometry of this system
are degenerate. Inspired by cosmological simulations that quantify structure
formation shocks \citep{2003ApJ...593..599R, 2008MNRAS.385.1211P,
  2008MNRAS.391.1511H}, we take the plausible assumptions that (1) the shock
surface is aligned with the gravitational equipotential surface of the Perseus
cluster that we determine by taking concentric spheres around NGC 1275 and (2)
the second outburst giving rise to the currently observed head-tail structure
has started shortly after shock crossing so that we can identify both events.

\subsection{Observables and Simple Derived Quantities}

\subsubsection{The Perseus accretion shock}  

NGC 1265 has a large radial infall velocity of $v_r=2170\,\rmn{km~s}^{-1}$ that
has been obtained after subtracting the mean heliocentric velocity of the
Perseus cluster of $v_{r,\,\rmn{Per}} = 5366\, \rmn{km~s}^{-1}$.  The Faraday RM
across the central few kpc of the two jets \citep{1986ApJ...301..841O} show a
scatter of $20~\rmn{rad\,m}^{-2}$ around a mean value of around
$25~\rmn{rad\,m}^{-2}$. The scatter is most probably due to the magnetized
plasma in the cocoon surrounding the jet or the interstellar medium of NGC 1265
\citep{1987ApJ...316...95O}. The low value of RM fluctuations strongly argues
for an infalling geometry with NGC 1265 sitting on the near side of Perseus and
little Faraday rotating material in between us and the source.  This argument
strongly favors the picture that the shock that transformed the radio bubble is
the accretion shock rather than a merger shock. The strongest argument in favor
of this picture would be the identification of the filament along which NGC 1265
was accreted in redshift space. Since NGC 1265 is only 600~kpc offset from the
center of the Perseus cluster (in projection), the finger-of-God effect makes it
very challenging to isolate a galaxy filament. There is however indirect
evidence for such a cosmic filament from Perseus X-ray data. Dark H$\alpha$
filaments near the LOS to the core do not show any foreground cluster
X-ray emission and have redshifted velocities of around 3000~$\rmn{km~s}^{-1}$ relative to
the Perseus cluster, so they appear to be falling in, too (A.~Fabian, private
communication).

\subsubsection{Electron cooling timescales}

The radio synchrotron radiating electrons of Lorentz factor $\Gamma$ emit at a
frequency
\begin{equation}
  \label{eq:nu_syn}
  \nu_\rmn{syn} = \frac{3 e B \Gamma^2}{2\pi m_e c},
\end{equation}
where $e$ is the elementary charge, $m_e$ is the electron mass, and $c$ is the
light speed.  Synchrotron and inverse Compton aging of relativistic electrons
occurs on a timescale
\begin{equation}
  \label{eq:tau_syn,ic}
  \tau_\rmn{syn,\,ic} = 
  \frac{6\pi m_e c}{\sigma_\rmn{T} (B_\rmn{cmb}^2+B^2) \Gamma},
\end{equation}
where $B_\rmn{cmb}\simeq3.2\mu\rmn{G}\,(1+z)^2$ and $\sigma_\rmn{T}$ is the
Thompson cross section. Combining both equations by eliminating the Lorentz
factor $\Gamma$ yields the cooling time of electrons that emit at frequency
$\nu_\rmn{syn}$,
\begin{eqnarray}
  \label{eq:taucool}
  \tau_\rmn{syn,\,ic} =
  \frac{\sqrt{54\pi m_e c\, e B \nu_\rmn{syn}^{-1}}}
  {\sigma_\rmn{T}\,(B_\rmn{cmb}^2+B^2)}
  \lesssim2.9 \times10^8 \,\rmn{yr}.
\end{eqnarray}
The highest frequency $\nu_\rmn{syn}=600\,\rmn{MHz}$ at which the torus can be observed
gives the shortest cooling time for any given magnetic field value.
Interestingly, $\tau_\rmn{syn,\,ic}$ is then bound from above and attains its
maximum cooling time at $B=B_\rmn{cmb}/\sqrt{3} \simeq
1.8\,\mu\rmn{G}\,(1+z)^2$.

\subsubsection{Limits on the galaxy's inclination angle} 

We denote the projected length on the plane of the sky that the galaxy has
traveled since shock crossing by $L_{t,\,\rmn{gal}}$. Looking at the bending of
the jet of NGC 1265 \citep{1986ApJ...301..841O} and the overall morphology of
the tail \citep{1998A&A...331..901S}, we conclude that the past orbit of NGC
1265 in the plane of the sky is directed mostly southward. We obtain a lower
limit on $L_{t,\,\rmn{gal}}\gtrsim 220$~kpc by measuring the vertical distance
of the ``head'' to the end of the bright radio tail (Figure~1(a) in
\citealt{1998A&A...331..901S}). Assuming that the time of shock passage of the
galaxy and the radio bubble coincide, we can obtain a lower limit on the
inclination of the galaxy's with the LOS,
\begin{equation}
  \label{eq:theta}
  \theta\geq
  \arctan\left[\frac{\rmn{min}(L_{t,\,\rmn{gal}})}
    {v_r\,\rmn{max}(\tau_\rmn{syn,\,ic})}\right]
  \simeq 19^\circ,
\end{equation}
where $v_r=2170\,\rmn{km~s}^{-1}$.  We will see that detailed geometric
considerations of the overall morphology of NGC 1265 show that there is a time
lag between the shock passage of the galaxy and the bubble of $\tau \simeq
6\times 10^7\,\rmn{yr}$ that one has to add to $\tau_\rmn{syn,\,ic}$ such that
the limit on the inclination weakens slightly and becomes $\theta \geq
16^\circ$.

\subsection{Geometrical Constraints}

It turns out that the morphological richness of NGC 1265 in combination with
physical arguments then constrain the geometry and shock properties of this
system surprisingly well. To this end, we derive a general expression for the
gas velocity after shock passage by taking into account the shock obliquity,
$\phi$, the shock orientation, $\chi$, and the inclination of the galaxy's
orbit, $\theta$ (see Table~\ref{tab:def} for definitions). Using mass and
momentum conservation at an oblique shock with a compression factor $C_s =
\rho_2/\rho_1$, the velocities parallel and perpendicular to the shock read as
\begin{eqnarray}
\label{eq:vshock}
\vecbf{v}_{2,\perp}    &=& \vecbf{v}_{1,\perp} = 
\vecbf{v} + v \cos\phi\, \vecbf{n}_s,\\
\vecbf{v}_{2,\parallel} &=& \frac{\vecbf{v}_{1,\parallel}}{C_s} =
-\frac{v}{C_s}\,\cos\phi\, \vecbf{n}_s,\\
\vecbf{v}_{2} &=&  \vecbf{v} + v \cos\phi\,\frac{C_s-1}{C_s}\, \vecbf{n}_s.
\end{eqnarray}
In order to connect to observables, we derive the velocity components parallel and
perpendicular to the LOS,
\begin{eqnarray}
  \label{eq:v2r}
v_{2,r} &=& \vecbf{e}_r\,\cdot\vecbf{v}_2
=v_r\,\left(1 - \frac{\cos\chi\cos\phi}{\cos\theta}\,\frac{C_s-1}{C_s}\right),\\
v_{2,t} &=& \sqrt{( \vecbf{v}_{2} - v_{2,r} \vecbf{e}_r)^2}
= \left[\frac{v_r^2}{\cos^{2}\theta}\left(1-\cos^2\phi\,\frac{C_s^2-1}{C_s^2} \right)\right.\nonumber\\
 &-& \left. v_r^2\,\left(1-\frac{\cos\chi\cos\phi}{\cos\theta}\,\frac{C_s-1}{C_s}\right)^2\right]^{1/2}.
 \label{eq:v2t}
\end{eqnarray}
We will determine the shock compression factor $C_s$ in Section~\ref{sec:shock}
based solely on the volume change of the bubble while assuming that the bubble
had negligible ellipticity prior to shock crossing.

We employ the following strategy to determine the geometry. (1) We choose the
accretion shock radius $R_s\gtrsim R_{200}=1.9\,$Mpc since the hot gas within
the cluster is unlikely to support a shock with Mach number $\M\simeq 4$. (2) A
priori, the position of the shock passage of NGC 1265 and its advected radio
bubble is unknown. We choose the point where the galaxy passed the shock as the
origin of the coordinate system $\mathcal{O}$, use the current galaxy's position
as a starting value, and interpolate the past orbits of the galaxy and the tail
in the plane of the sky. An iteration of the position of $\mathcal{O}$ (that
involves a complete cycle through the points listed here) quickly converges on a
consistent model. Subject to our assumption of the shock surface, this yields
the shock normal, $\vecbf{n}_s$, the shock orientation, $\chi$, and projected
orbit that the galaxy has traveled since shock crossing, $L_{t,\,\rmn{gal}}$.
(3) Choosing the inclination of the galaxy's orbit, $\theta$, determines the 3D
velocity vector $\vecbf{v}$ and the time since shock crossing,
$\tau_s=L_{t,\,\rmn{gal}}/ (v_r \tan\theta)$. (4) Momentum conservation at the
shock implies for the post-shock gas that its velocity vector lies in the plane
that contains $\vecbf{n}_s$ and $\vecbf{v}$ and is defined by
$\vecbf{r}\cdot(\vecbf{n}_s\times \vecbf{v}/|v|)=0$. The intersection of this
plane and the plane of the sky (given by $z=0$ for our choice of coordinates)
yields the past transverse orbit of the post-shock gas, $v_\rmn{2,t}$. (5)
Solving Equation~(\ref{eq:v2t}) for the unknown shock obliquity $\phi$ enables
us to derive the current radial position of the galaxy (relative to the cluster
center) according to $r_\rmn{gal}=R_s - v \tau_s \cos\phi$ with
$v=v_r\sqrt{1+\tan^2\theta}$.  (6) We compare the galaxy's velocity, $v$, to the
escape velocity, $v_\rmn{esc}(r_\rmn{gal})=\sqrt{2G
  M(<r_\rmn{gal})/r_\rmn{gal}}$. For the large radial velocity of
$v_r=2170\,\rmn{km~s}^{-1}$, it is not trivial to meet the criterion, $v\lesssim
v_\rmn{esc}$, stating that the galaxy is gravitationally bound; hence we prefer
small values of $R_s$ and $\theta \lesssim 32^\circ$ (while still evading the
electron cooling bound on $\theta$ as derived in Equation~(\ref{eq:theta})). The
head-tail morphology of the jet also argues for a large ram pressure that it can
only experience inside $\sim R_{200}$ and hence for small values of $R_s$. We
finally check whether the model violates any constraints on the shock obliquity
$\phi$ as will be derived in Section~\ref{sec:stability}. If this does not yield a
consistent model we vary the shock radius $R_s$, the inclination of the galaxy's
orbit $\theta$, and the position of the galaxy's shock crossing, $\mathcal{O}$,
and start the next iteration until we arrive at a self-consistent and physically
plausible model.

\begin{figure}
\epsscale{0.95}
\plotone{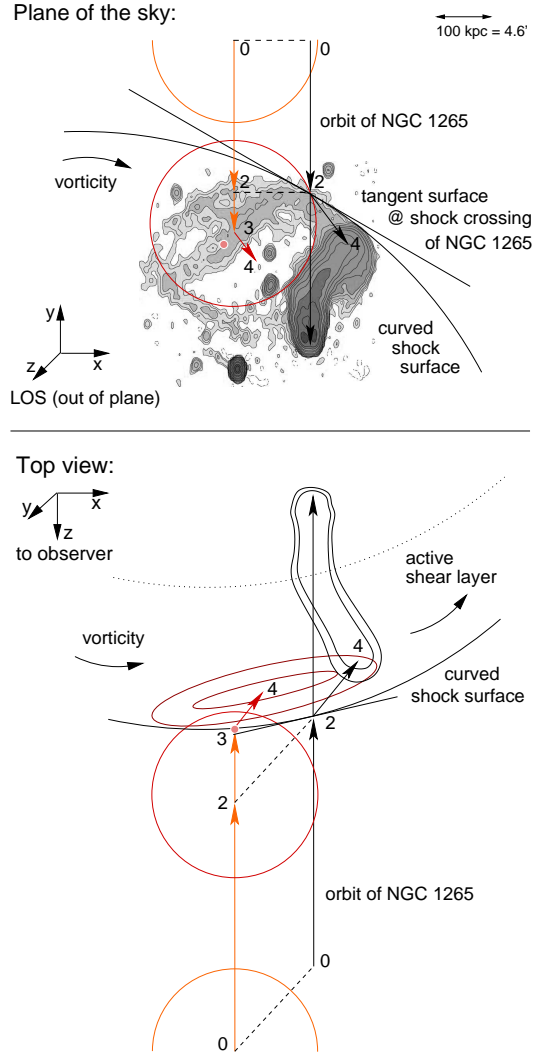}
\caption{Plane of the sky view (top) and view from above (bottom) of
  the time evolution of NGC 1265 through the accretion shock of Perseus
  \citep[radio surface brightness at 600~MHz taken
  from][]{1998A&A...331..901S}. Note that the angles of the cut-planes through
  the shock surfaces (at the point of shock passage of NGC 1265, labeled with 2)
  and velocity vectors are derived from our best-fitting geometry, are drawn to
  scale (given at the top right), and the numbers correspond to the time
  sequence defined in Figure~\ref{fig1}.  The galaxy and the detached radio bubble
  are advected in a filament (initial configuration labeled with 0, bubble on
  the left shown in projection as orange half-circle, bubble on the right in
  Figure~\ref{fig1} omitted for clarity).  The galaxy stays on its ballistic orbit
  unaffected by the shock passage (2). In the top panel, the shock surface is
  almost face-on with a small inclination of $\chi=23\degr$ to the LOS
  whereas the bubble is offset from the cut plane and situated closer to the
  observer. This inclination of the jet axis from the plane of the sky implies a
  later shock crossing time of the bubble compared to the galaxy.  After
  $\tau_{2\to3}\simeq 6\times 10^7\,\rmn{yr}$, the bubble passes through the
  shock and transforms into a torus (3). The radio torus and the tail of NGC
  1265 experience an almost coherent deflection due to the oblique shock passage
  for the remaining time of $\tau_{3\to4}\simeq 1.2\times 10^8\,\rmn{yr}$ until
  today (4). The red circle shows the bubble circumference when the shock
  surface comes into contact with the bubble (3, shown as light red point); due
  to the inclination of the shock with respect to the LOS, this point would not
  be visible in projection as it appears in both cases at the back side of
  the bubble. The fact that projected shock surface coincides with the rim of
  the torus is accidental and depends on the choice of the radial position of
  the shock $R_s$ which we choose as $R_s\simeq R_{200}$.  A shell of 250~kpc in
  the region interior to the accretion shock appears to experience an additional
  shear flow caused by vorticity generation at the curved shock surface.  }
  \label{fig3}
\end{figure}

Our final model that fits best these constraints is shown in Figure~\ref{fig3}
and has the parameters $C_s=3.4$, $R_s=R_{200}=1.9\,$Mpc, $\theta=32^\circ$,
$\phi=9^\circ$, $\chi=23^\circ$.  We find the time since shock crossing of NGC
1265 to be $1.8\times10^8\,\rmn{yr}$, the Cartesian vector of the shock normal
$\vecbf{n}=(0.2,0.34,0.92)$, and the velocity vector for NGC 1265 of
$\vecbf{v}=-v\,(0,\sin\theta,\cos\theta)$ (note that we define the Cartesian
coordinate system in Figure~\ref{fig3}). The galaxy's velocity has a radial and
transverse component of $v_r=2170\,\rmn{km~s}^{-1}$ and
$v_t=1360\,\rmn{km~s}^{-1}$, respectively, which yields a total velocity of the
galaxy, $v=2550\,\rmn{km~s}^{-1}$. This is only slightly larger than the escape
velocity, $v_\rmn{esc}(r_\rmn{gal}) = 2350\,\rmn{km~s}^{-1}$, where we used
$r_\rmn{gal}=1.45\,$Mpc, $M(<r_\rmn{gal})\simeq X_\rmn{turb}
M_{200}=9\times10^{14}M_\odot$, neglect a logarithmic correction factor of the
mass and assume a turbulent pressure support of $X_\rmn{turb}=0.2$
\citep{2008Sci...320..909R, 2009ApJ...705.1129L, 2010ApJ...725...91B}. Note that
if one relaxes the escape velocity constraint and assumes that tidal processes
are able to dissipate more energy to bind the galaxy during first passage, we
find consistent solutions at larger shock radii that nevertheless lie close-by
in the parameter space $(\theta,\phi,\chi,\mathcal{O})$. The choice of the
virial radius as the site of the accretion shock might seem to be too small
compared to the typical locations inferred from cosmological simulations
\citep{2000ApJ...542..608M, 2007ApJ...669..729K, 2008MNRAS.385.1211P}. However,
the location of the accretion shock (in particular along a filament) is not
stationary but dynamically determined. Depending on the ram pressure of the
accreting material and the post-shock pressure, its position will re-adjust
dynamically to account for the conservation laws.  We have shown that the
velocity of NGC 1265 is quite large which implies a large ram pressure and hence
should yield a shock position that lies closer toward the cluster center
compared to the shock position where dilute IGM accretes from voids with a
considerable smaller density and ram pressure.  We note that there have been
attempts to constrain the 3D velocity of NGC 1265 based on the observed
sharpness of its X-ray surface brightness map and the optical velocity
dispersion of the Perseus cluster \citep{2005ApJ...633..165S}. To arrive at a
lower bound on the orbit inclination, $\theta_\rmn{X-ray}>45\degr$, these
authors use an axially symmetric density profile that however does not take into
account magnetic draping. This naturally would account for the sharpened
interface between the ICM and the interstellar medium and might break the
assumed symmetry depending on the morphology of the ambient magnetic field
\citep{2008ApJ...677..993D, 2010NatPh...6..520P}.

The post-shock gas velocity---as traced by the radio emitting tail---is
reduced at the shock by transferring a fraction of the kinetic to internal
energy. For our choice of parameters, we find
$v_2=v\sqrt{1-\cos^2\phi\,(C_s^2-1)/C_s^2}\simeq850\,\rmn{km~s}^{-1}$ that divides
into the radial and transverse LOS component of the tail as
$v_{2,r}\simeq520\,\rmn{km~s}^{-1}$ and $v_{2,t}\simeq670\,\rmn{km~s}^{-1}$, respectively.
The velocity components parallel and perpendicular to the shock normal are
$v_{1,\parallel}=v\cos\phi\simeq 2530\,\rmn{km~s}^{-1}$,
$v_{2,\parallel}=v\cos\phi/C_s\simeq 740\,\rmn{km~s}^{-1}$, and
$v_\perp=v_{1,\perp}=v_{2,\perp}=v\sin\phi\simeq 400\,\rmn{km~s}^{-1}$.  Once in the
post-shock regime, the radio emitting tail and the torus experience the same
large-scale velocity field. To explain the observed direction of the bending of
the tail toward west, the shock normal needs to have a component pointing
westward (for the observed direction of the galaxy's velocity southward).

What about the orientation of the line connecting the bubble center and NGC 1265
(shown as dashed lines in Figure~\ref{fig3})?  Assuming that the spin orientation
of the SMBH as possibly traced by the radio jets did not change during shock
passage, the previous jet blowing the radio bubbles predominantly had an east-west (EW)
component in the plane of the sky. If the jet angle had no component along the
LOS, the bubble had crossed the shock earlier than NGC 1265 due to the assumed
convex shock curvature. Hence the bubble would need to have been at a larger
distance from NGC 1265 than in our model and would have needed a longer time
$\tau\simeq 3\times 10^8\,\rmn{yr}>\tau_\rmn{syn,\,ic}$ to reach the current
position (extrapolating the position back to the shock with the orientation of
the deflection vector given by our model). It follows that the shock-enhanced
radio emission would have faded away by now and the bubble would have had to
spend at least a factor of 2.5 longer in the post-shock region which would not
be consistent with the assumed EW alignment of the previous jet axis. Hence our
model postulates that the eastern bubble was on the front side relative to the
observer and had a smaller LOS than NGC 1265 (see Figure~\ref{fig3}). This caused
a ``two-stage process'': the bubble continued for $\tau_{2\to3}\simeq 6\times
10^7\,\rmn{yr}$ on its orbit in the filament after NGC 1265 passed the shock
until it came into contact with the shock surface. This configuration has the
advantage of the faster pre-shock velocity and associated larger length scale in
the plane of the sky that the bubble traversed and enables us to fulfill the
geometric requirement to cross the shock surface while ending up at current
position.  After shock passage and its transformation into a torus, it
experiences the an almost similar deflection as the tail of NGC 1265 due to the
oblique shock passage for the remaining time of $\tau_{3\to4}\simeq 1.2\times
10^8\,\rmn{yr}$ until today.  This also solves the problem of the
non-observation of the ``second'' bubble that the previous western jet should have
blown: in our proposed geometry, this hypothetical bubble was situated westward
of NGC 1265 at the far side of the cut plane in Figure~\ref{fig3}. It should have
passed the shock $\tau\sim 2\times 10^8\,\rmn{yr}$ earlier than the eastern
bubble and faded away by now (Equation~(\ref{eq:taucool})). Note that the position
and morphology of the accretion shock is a function of time: as the radio bubble
comes into contact with the shock surface, the shock expands into the light
radio plasma toward the observer and attains additional curvature. For
simplicity we suppress this effect in our sketch in Figure~\ref{fig3} and place
the torus in the lower panel at a distance from the (initial and unmodified)
shock surface which should be equal to that of the modified shock surface.


\section{Properties of the Perseus Accretion Shock}
\label{sec:shock}

Using the 49 cm image of NGC 1265 (Figure~1(a) in \citealt{1998A&A...331..901S}), we
measure the major and minor radius of the torus to $R\simeq7\arcmin$ and
$r_\rmn{min}\simeq 1.\arcmin1-1.\arcmin4$. We only consider orientations that
lie in the northern and southern sectors as the elongated appearance toward the
EW might be caused by shear flows as we will argue later on in this
section. At the distance of Perseus, this corresponds to physical length scales
of $R \simeq 150\,\rmn{kpc}$ and $r_\rmn{min} \simeq (24\ldots30)\,\rmn{kpc}$,
respectively.  Assuming that the bubble was spherical before shock crossing and
that the major radius did not change \citep{2002MNRAS.331.1011E}, we estimate a
compression factor through the associated volume change of
\begin{equation}
  \label{eq:compression}
 C=\frac{V_\rmn{bubble}}{V_\rmn{torus}} = 
 \frac{\frac{4}{3}\pi R^3}{2 \pi^2 R r_\rmn{min}^2} = 
\frac{2}{3\pi}\,\left(\frac{R}{r_\rmn{min}}\right)^2\simeq 6-10. 
\end{equation}
The radio plasma is adiabatically compressed across the shock passage according
to $P_2 / P_1 = C^{\gamma_\rmn{rel}}$, where $\gamma_\rmn{rel} = 4/3$ for an
ultra-relativistic equation of state. However, the adiabatic index can in
principle approach $\gamma_\rmn{rel} \lesssim 5/3$ if the bubble pressure is
dominated by a hot but non-relativistic ionic component. Varying
$\gamma_\rmn{rel}\simeq1.33-1.5$ and $C\simeq 6-10$, we obtain
$P_2/P_1\simeq21.5\pm10.5$. Assuming that the radio bubble is in pressure
equilibrium with its surroundings before and after shock crossing, this pressure
jump corresponds to the jump at the shock.  Applying standard Rankine Hugoniot
jump conditions \citep{1959flme.book.....L} of an ideal fluid of adiabatic index
$\gamma=5/3$ yields a shock Mach number of $\M\simeq 4.2_{-1.2}^{+0.8}$, a
density jump of $3.4_{-0.4}^{+0.2}$, and a temperature jump of
$6.3_{-2.7}^{+2.5}$.  We assumed a flat prior on the uncertainties of $C$ and
$\gamma_\rmn{rel}$ which yields symmetric error bars around the pressure jump.
We quote the jumps in the other thermodynamic quantities corresponding to our
mean $P_2/P_1$. These quantities inherit asymmetric error bars due to the
non-linear dependence of these jumps on Mach number.  We caution that we
performed these estimates from the 600 MHz map \citep{1998A&A...331..901S};
future higher resolution and more sensitive radio observations are needed to
assess the uncertainties associated with finite beam width and to obtain a more
reliable morphology of the low surface brightness regions of the torus.

Extrapolating X-ray profiles of Perseus \citep{2003ApJ...590..225C} to
$R_{200}=1.9$~Mpc and using our mean values for the jumps in thermodynamic
quantities, we can derive pre-shock values for the gas temperature and density
that reflect upper limits on the gas properties in the infalling warm-hot IGM as
predicted by \citet{2009MNRAS.393.1073B}. We find $kT_1\lesssim0.4$~keV, $n_1 =
1.93\, n_{1,e}\lesssim 5\times 10^{-5}\,\rmn{cm}^{-3}$, and $P_1\lesssim
3.6\times 10^{-14}\, \rmn{erg\,cm}^{-3}$. For the cluster temperature profile,
we combine the central profile from X-ray data with the temperature profiles
toward the cluster periphery according to cosmological cluster simulations
\citep[e.g.,][]{2007MNRAS.378..385P, 2010MNRAS.409..449P}, yielding
\begin{equation}
  \label{eq:temp}
  kT(r) = \left[kT_0 + \frac{kT_\rmn{max} - kT_0}{1 + (r/r_{c})^{-3}}\right]\,
 \left[1 + \left(\frac{r}{0.2\,R_{200}}\right)^2\right]^{-0.3},
\end{equation}
where $kT_0=3\,\rmn{keV}$, $kT_\rmn{max}=7\,\rmn{keV}$, and $r_{c}=94\,\rmn{kpc}$.  This
represents the correct temperature profile over the entire range of the cluster
when comparing to mosaiced observations by Suzaku (S. Allen, private
communication).

We now present evidence for post-shock shear flows with an argument that is
based on the projected orientation of the ellipsoidally shaped radio torus whose
main axis is almost aligned with the EW direction. We consider two cases. (1) If
the ellipticity was due to projection of a ring-like torus, the shock normal
$\vecbf{n}_s$ could not have components in the EW direction. Since the
transverse velocity of the galaxy is pointing southward, momentum conservation
at the (oblique) shock would imply a deflection of the post-shock gas in the
plane that contains $\vecbf{n}_s$ and $\vecbf{v}$---without components in the EW
direction. Hence an additional shear flow would be needed to explain the
westward bending of the galaxy's tail even though the smooth and coherent
bending of the tail might be difficult to reconcile with the turbulent nature of
vorticity-induced shear flows.  (2) If the bending of the tail was due to
oblique shock deflection, $\vecbf{n}_s$ would have to have a component pointing
westward. Projecting an intrinsically ring-like torus---aligned with the shock
surface---would yield an apparent ellipsoidal torus with the main axis at some
angle with the EW direction on the plane of the sky. This argues for a shear
flow that realigns the orientation of the ellipsoid with the observed EW
direction.  In fact, assuming that the shock surface is aligned with the
gravitational equipotential surface of the Perseus cluster, one can show that
the implied curvature of the shock surface causes a vorticity in the post-shock
regime that shears the post-shock gas westward\footnote{The sign of the
  post-shock vorticity is given by the baroclinic term, $\nabla\rho\times\nabla
  P$, which in the top view of Figure 1 (lower panel) points out of the page.}.
We cannot completely exclude pre-existing ellipticity in the radio bubble but
find it very unlikely to be the dominant source of the observed ellipticity as
the same directional deflection of the tail structure as well as the radio torus
would then be a pure coincidence rather than a prediction of the model.

The observed flattening of the projected radio torus amounts to $f=1-b/a\simeq
0.3$, where $a$ and $b$ are the major and minor axes of the ellipse,
respectively. With the assumption that the shock surface is aligned with the
gravitational equipotential surface and using the result from
Section~\ref{sec:geometry}, the flattening due to projection,
$f=1-\cos\phi\simeq0.066$, falls short of the observed one calling again for
shear flows to account for the difference.  The amount of vorticity $\omega$
injected at a curved shock of curvature radius $R_\rmn{curv}$ should be
proportional to the perpendicular velocity $v_\perp$ over the bubble radius
$r_\rmn{bubble}$ \citep{1957JFM.....2....1L},
\begin{equation}
  \label{eq:vorticity2}
  \omega = \frac{(C_s-1)^2}{C_s} \frac{v_\perp}{R_\rmn{curv}} = 
  \tilde{f} \frac{v_\perp}{r_\rmn{bubble}},
\end{equation}
where the constant of proportionality $\tilde{f}$ depends on the flattening $f$ of
the torus. Using $\tilde{f}\simeq f\simeq0.3$, $C_s\simeq3.4$, and
$r_\rmn{bubble}=150$~kpc, we obtain a rough estimate for the curvature radius of
the shock of $R_\rmn{curv}\simeq850$~kpc which is also adopted in
Figure~\ref{fig3}. This is similar to the dimensions of a filament connecting to a
galaxy cluster and provides an independent cross-check of our picture.

The ratio of the energy density in the shock-injected vorticity (or shear flow)
to the thermal energy density is given by
\begin{equation}
  \label{eq:turb3}
  \frac{\eps_\rmn{shear}}{\eps_\rmn{th,2}}
  = \frac{\mu m_p v_{\perp}^2}{3 kT_2}\simeq 0.14,
\end{equation}
where $kT_2\simeq 2.4\,\rmn{keV}$ and $v_\perp\simeq
400\,\rmn{km~s}^{-1}$. Interestingly, this value is very close to the
turbulent-to-thermal pressure support of $\eps_\rmn{turb}/\eps_\rmn{th<}\simeq
0.2$ at $R_{200}$ found in cosmological simulations of the formation of galaxy
cluster \citep{2008Sci...320..909R, 2009ApJ...705.1129L, 2010ApJ...725...91B}.
We caution that our estimates strongly depend on our model assumptions for the
orientation and position of the shock surface with respect to the LOS since
projection effects of a circular torus could account for a fraction of the
observed ellipticity, reduce the shear contribution, and increase the inferred
curvature radius. Another complication is the amount of magnetic helicity inside
the torus that acts as a stabilizing agent against the shear that tries to
distort and potentially disrupts the plasma torus \citep{2007MNRAS.378..662R,
  2010MNRAS.406..705B}.


\section{Scrutinising our model of NGC 1265}
\label{sec:model}

\subsection{Examining the Physics of the Model in Detail}
\label{sec:detail}

Using results on the shock properties and geometric parameters from
Sects.~\ref{sec:geometry} and \ref{sec:shock}, we demonstrate the physical
feasibility of this model.

\subsubsection{The radio bubble and torus}

The expansion time of the bubble in the filament (prior to shock passage) is of
order the external sound crossing time, $\tau_\rmn{exp}= R/c_\rmn{1}\simeq
0.4\,\rmn{Gyr}$, using a a sound speed of the pre-shock region $c_\rmn{1}\simeq
285\,\rmn{km~s}^{-1}\,(kT/0.4 \,\rmn{keV})^{1/2}$ and a bubble radius of
$R\simeq 150\,\rmn{kpc}$. This assumes that the outer radius $R$ is unchanged
upon transformation into a torus (as suggested by numerical simulations of
\citealt{2002MNRAS.331.1011E}) and that $R$ equals the minor axis of the ellipse
of the projected torus that is unaffected by the shear. We can estimate the
energetics of the previous outflow by calculating the $P\dd V$ work done in
inflating the bubble. Using the pre-shock gas pressure of $P_1\simeq
3.6\times10^{-14}\,\rmn{erg~cm}^{-3}$, we obtain $E_\rmn{bubble} =
1/(\gamma_\rmn{rel}-1) P_1 V\simeq 4\times 10^{58}\,\rmn{erg}$ using
$\gamma_\rmn{rel}=4/3$. This fits well inside the range of observed jet energies
of FRI sources and implies an equipartition magnetic field in the bubble of
$B_\rmn{1,eq} = (8\pi P_1/2)^{1/2}\simeq 0.66\,\mu$G that becomes after shock
passage $B_{2,\rmn{eq}}=B_{1,\rmn{eq}}\,C^{2/3}\simeq 3\,\mu$G with $C\simeq10$;
somewhat higher than $B_\rmn{eq}\simeq1\,\mu$G obtained by equipartition
arguments applied to the radio flux \citep{1998A&A...331..901S}.  The radio
synchrotron radiating electrons of Lorentz factor $\Gamma$ emit at a frequency
$\nu_\rmn{syn}$ given by Equation~(\ref{eq:nu_syn}).  Using a range of
magnetic field values of $B_\rmn{eq}\simeq(1-3)\,\mu\rmn{G}$, we require
electrons with a post-shock Lorentz factor $\Gamma_2\simeq 4800-8500$ to explain
the observed radio torus at 600~MHz and derive a minimum value of
$\Gamma_1=\Gamma_2/C^{1/3}\simeq 2200-4000$ for the upper cutoff of the electron
population in order to be able to account for the observed radio emission after
shock passage.  Synchrotron and inverse Compton aging of fossil electrons in the
bubble occurs on a timescale
\begin{equation}
  \label{eq:tau_syn,ic2}
 \tau_\rmn{syn,\,ic, 1} = 
 \frac{6\pi m_e c}{\sigma_\rmn{T} (B_\rmn{cmb}^2+B_\rmn{1,eq}^2) \Gamma_1} = 
 (0.6-1) \,\rmn{Gyr} > \tau_\rmn{exp}.
\end{equation}
Calculating the maximum cooling time that the bubble could have been hibernating
as given by Equation~(\ref{eq:tau_syn,ic2}), we show that we meet this criterion for
$\Gamma_1$ so that the model by \citet{2001A&A...366...26E} applies.

\subsubsection{The current outburst and radio tail of NGC 1265} 

Assuming that the twin tails together constitute a cylinder of radius
$r_\rmn{tail}\simeq1.\arcmin5$ and projected length
$L_{t,\,\rmn{tail}}\simeq11\arcmin$, we can estimate the volume for the radio
tail of NGC 1265, $V_\rmn{tail}\simeq\pi r_\rmn{tail}^2
L_{t,\,\rmn{tail}}/\tan\theta\simeq 3.8\times 10^{70}\,\rmn{cm}^3$ using
$\theta\simeq 32\degr$.  Taking the minimum energy density from
\citet{1998A&A...331..901S} in the tail of NGC 1265 to be around
$\eps_\rmn{eq}\simeq 10^{-12} \,\rmn{erg~cm}^{-3}$ we estimate the energy content
of the radio tail to be about $E_\rmn{jet}\simeq3.8\times 10^{58} \,\rmn{erg}$
which nicely coincides with the result that we derived for the previous
outburst. The necessary accreted mass to power the outflow amounts to
$M_\rmn{acc}= E_\rmn{jet}/(\eta\,c^2)\simeq 2\times 10^5 \,M_\odot$, where we
adopted a typical values of $\eta\simeq0.1$ for the efficiency parameter.  Using
the time since shock crossing in our model of $\tau\simeq 1.8\times
10^8\,\rmn{yr}$, we estimate the jet power of about $6.7\times
10^{42}\,\rmn{erg~s}^{-1}$ which is in range of observed jet luminosities of FRI
sources. Using the 1.4 GHz flux for NGC 1265 of 8~Jy, we obtain a radio
luminosity of $\nu L_{\nu} \simeq 7.5\times10^{40} \,\rmn{erg~s}^{-1}$, or a 1\%
radiative efficiency which is plausible.

\subsubsection{Estimates from the jet bending} 

Using the high-resolution VLA observations of the head structure of NGC 1265,
\citet{1986ApJ...301..841O} give a minimum jet pressure $P_\rmn{jet,\,min}\simeq
(1-3)\times 10^{-11}\,\rmn{erg\,cm}^{-3}$ which is about 10 times the ICM
estimate at the present position of the galaxy, $P_\rmn{ICM}(r_\rmn{gal}) \simeq
1.4\times 10^{-12}\, \rmn{erg\,cm}^{-3}$ with $r_\rmn{gal}\simeq1.45\,\rmn{Mpc}$
(see Sects.~\ref{sec:geometry} and \ref{sec:shock}). The jet radius is
$r_\rmn{jet}\lesssim 1\arcsec$ (360 pc), whereas the projected bending radius
of the jet is around $30\arcsec$ (11 kpc), so the physical ratio is $r_b
/r_\rmn{jet} \sim 80$ when accounting for deprojection and finite resolution
effects. Using Equation~(\ref{eq:bending2}), this leads to a Mach number ratio
of $\M_\rmn{jet}/\M_\rmn{gal} \simeq 2.8$. The temperature estimate at the
position of the galaxy is $kT(r_\rmn{gal}) \simeq 3.1 \,\rmn{keV}$, so that we
obtain $\M_\rmn{gal} \simeq 2.8$ and eventually $\M_\rmn{jet} \simeq 7.8$. If we
take the jet power to be $L_\rmn{jet}/2 \sim \rho_\rmn{jet} v_\rmn{jet}^3 \pi
r_\rmn{jet}^2$ (where $L_\rmn{jet}$ is the luminosity of the 2 jets and we
assume that most of the jet power is kinetic energy), the power can be expressed
in terms of the jet pressure and Mach number as
\begin{equation}
  \label{eq:jet_power}
  \frac{L_\rmn{jet}}{2} \simeq \gamma_\rmn{jet} P_\rmn{jet} \M_\rmn{jet}^2 v_\rmn{jet}
  \pi r_\rmn{jet}^2.
\end{equation}
Since our estimate of $L_\rmn{jet} \simeq 6.7 \times 10^{42} \,\rmn{erg~s}^{-1}$,
we get $v_\rmn{jet} \simeq 7000 \,\rmn{km~s}^{-1}$ which is a typical jet
velocity. Solving for the number density of the jet material, we get
$n_\rmn{jet}\simeq L_\rmn{jet}/(2\pi m_p v_\rmn{jet}^3 r_\rmn{jet}^2)\simeq
10^{-3}\,\rmn{cm}^{-3}$ which is a factor of three larger than the surrounding
ICM, $n(r_\rmn{gal}) \simeq 3\times 10^{-4} \,\rmn{cm}^{-3}$ and again a
plausible value.

\subsection{Stability Analysis of the Torus}
\label{sec:stability}

There are three mechanisms that have the potential to destroy the torus once it
has formed; namely, (1) large scale shear flows injected at the shock front, (2)
ICM turbulence, and (3) Kelvin-Helmholtz instabilities at the interface due to
the vortex flow around the minor circle of the torus. By considering each of
these processes separately, we will show that none of them can seriously impact
the torus on an eddy turnover timescale of the stabilizing vortex flow,
\begin{equation}
  \label{eq:eddy}
  \tau_\rmn{eddy}\simeq \frac{2\pi r_\rmn{min}}{v_{2,\parallel}}
  \simeq 2\times 10^8\,\rmn{yr},
\end{equation}
where $r_\rmn{min}\simeq25\,\rmn{kpc}$ denotes the minor circle of the torus,
$v_{2,\parallel} = v\cos\phi / C_s\simeq 740\,\rmn{km~s}^{-1}$ is the
post-shock gas velocity parallel to the shock normal, $v\simeq2550\,\rmn{km~s}^{-1}$
the total velocity of the galaxy, and $\phi\simeq9\degr$ is the shock obliquity
defined in Table~\ref{tab:def}.  Here and in the following, we apply the
numerical values of our consistent model as derived in Sects.~\ref{sec:geometry}
and \ref{sec:shock}. On a timescale larger than $\tau_\rmn{eddy}$, dissipative
effects will thermalize the kinetic energy of this stabilizing flow, and
eventually external shear flows and turbulence might start to destroy the torus
depending on the magnetic helicity and morphology in the torus
\citep{2007MNRAS.378..662R, 2010MNRAS.406..705B}.

(1) The amount of vorticity $\omega_\rmn{shear}$ injected at a curved shock with
curvature radius $R_\rmn{curv}$ is given by \citep{1957JFM.....2....1L}
\begin{equation}
  \label{eq:vorticity}
  \omega_\rmn{shear} = \frac{(C_s-1)^2}{C_s} \frac{v_{\perp}}{R_\rmn{curv}}
  = \frac{(C_s-1)^2}{C_s} \frac{v \sin\phi}{R_\rmn{curv}},
\end{equation}
where $C_s$ is the shock compression factor and $v_{\perp}$ is the
(post-)shock gas velocity perpendicular to the shock normal. Note that for
vorticity injection into an irrotational flow encountering a shock, one
necessarily needs a curved shock surface according to Crocco's theorem
\citeyearpar{Crocco}. The generation of vorticity relies on a gradient of either
the entropy or stagnation enthalpy along the shock front (perpendicular to the
shock normal).  This can be easiest seen by considering a homogeneous flow
encountering the shock surface at some constant angle (obliquity). It
experiences the same ``shock deflection'' and amount of entropy injection along
the shock front. This is because mass and momentum conservation across the shock
ensures the conservation of $v_{\perp}$ and yields $v_{2,\parallel} =
v_{1,\parallel} / C_s$.  If the bubble was impacting the shock surface at
larger angle, the induced stabilizing vortex flow around the minor circle of
radius $r_\rmn{min}$ is accordingly smaller since the perpendicular velocity
component cannot contribute to the vortex flow,
\begin{equation}
  \label{eq:torus}
  \omega_\rmn{torus} = \frac{v_{2,\parallel}}{r_\rmn{min}}
  = \frac{v\cos\phi}{C_s r_\rmn{min}}.
\end{equation}
We can then derive a criterion for the stability of the torus in the presence of
shock injected shear flows, $\omega_\rmn{shear} < \omega_\rmn{torus}$ that can
be cast into a requirement for the shock obliquity,
\begin{equation}
  \label{eq:shear_phi}
  \phi < \phi_\rmn{crit} = 
  \arctan\left[\frac{R_\rmn{curv}}{r_\rmn{min} (C_s - 1)^2} \right]
  \simeq 80\degr.
\end{equation}
Here $C_s\simeq3.4$, $R_\rmn{curv} \simeq 850\,\rmn{kpc}$ as inferred from
the observed ellipticity of the torus, and $r_\rmn{min}\simeq 25
\,\rmn{kpc}$. Since our preferred value of the shock obliquity
$\phi=9\degr\ll\phi_\rmn{crit}$, the torus can easily be stabilized by the
vortex flow around its minor circle against large-scale shear.

(2) To assess the impact of ICM turbulence on the torus' stability, we compare
the turbulent energy density $\eps_\rmn{turb}$ to the kinetic energy density of
the torus vortex flow $\eps_\rmn{vort}$,
\begin{eqnarray}
  \label{eq:turb1}
  \left.\frac{\eps_\rmn{turb}}{\eps_\rmn{vort}}\right|_{r_\rmn{min}}&=& 
  \left.\frac{\eps_\rmn{turb}}{\eps_\rmn{th, 2}}\right|_{r_\rmn{min}} 
  \frac{\eps_\rmn{th, 2}}{\eps_\rmn{vort}} \nonumber\\
  &=& \frac{\eps_\rmn{turb}}{\eps_\rmn{th, 2}}
  \left(\frac{R_\rmn{curv}}{r_\rmn{min}}\right)^{-2/3}
  \frac{3 kT_2}{\mu m_p v_{2,\parallel}^2}\simeq 0.04.
\end{eqnarray}
Here, $\eps_\rmn{th, 2}$ and $kT_2=2.4\,\rmn{keV}$ denote the post-shock energy
density and internal energy, $\mu=0.588$ denotes the mean molecular weight for a
medium of primordial element abundance, and $m_p $ denotes the proton rest mass.
To quantify the ratio $\eps_\rmn{turb}/\eps_\rmn{vort}$ in our model, we assume
a turbulent pressure support relative to the thermal pressure at $R_{200}$ of
around 20\% when correcting for the energy contribution due to bulk flows
\citep{2008Sci...320..909R, 2009ApJ...705.1129L, 2010ApJ...725...91B}. We
furthermore assume that the turbulent energy density is dominated by the
injection scale (which should be comparable to the curvature radius at the
shock). Using a Kolmogorov scaling of the turbulence, the power per logarithmic
interval in scale length is down by another factor of
$(R_\rmn{curv}/r_\rmn{min})^{2/3}\simeq 10$ at the minor radius $r_\rmn{min}$
compared to the injection scale $R_\rmn{curv}$.  Hence, the energy density of
the stabilizing vorticity flow around the minor axis of the torus outweighs any
turbulent energy density of the ICM on this scale by a factor of about 25!
Equation~(\ref{eq:turb1}) enables us to solve for the maximum shock obliquity
allowed so that the torus is stable against turbulent shear flows,
\begin{equation}
  \label{eq:turb2}
  \phi < \arccos\left[
 \frac{\eps_\rmn{turb}}{\eps_\rmn{th}}\left(\frac{R_\rmn{curv}}{r_\rmn{min}}\right)^{-2/3}
  \frac{3 kT_2 C_s^2}{\mu m_p  v^2} \right]^{1/2}\simeq 78\degr,
\end{equation}
where $v\simeq2550\,\rmn{km~s}^{-1}$ and $C_s\simeq 3.4$. Interestingly, this
is a similar value as $\phi_\rmn{crit}$ obtained in Equation~(\ref{eq:shear_phi}) and
is another manifestation of our finding that $\eps_\rmn{turb}\sim
\eps_\rmn{shear}$ (Equation~(\ref{eq:turb3})) implying that a large fraction of
post-shock turbulence could have been injected at curved shocks.

(3) We finally show, that the Kelvin-Helmholtz instability at the interface of
the torus due to the vortex flow should be suppressed by internal magnetic
fields.  The shock compression acts mostly perpendicular to the forming toroidal
ring. Consequently, the field component parallel to the bubble surface and hence
to the surface of the torus is preferentially amplified and dominates eventually
the magnetic energy density \citep{2002MNRAS.331.1011E}.  Within the torus, we
assume equipartition magnetic fields of $B_\rmn{eq,2}\simeq (1$--$3)\mu\rmn{G}$
which correspond to radio and pressure equipartition values, respectively
(Section~\ref{sec:detail}).\footnote{If the magnetic field was sub-equipartition
  in some regions just below the torus surface, the external vortex flow implies
  a matching internal vortex flow that acts in piling up the field in a layer
  just internal to the surface. In principle internal vortex flows could be on
  small scales as long as the effective internal vorticity matches the external
  one. However buoyancy of the magnetic field in combination with an effective
  reshuffling of the magnetic field due to internal vortex flows should give
  rise to a surface layer filling magnetic field.}  Assuming the torus to be in
pressure equilibrium with the surroundings and a relativistic electron component
in the bubble whose pressure is dominated by momenta $\sim m_e c$, we can derive
a density contrast between the radio plasma and post-shock density of the ICM,
\begin{equation}
  \label{eq:delta}
  \delta = \frac{n_\rmn{torus}}{n_\rmn{2, ICM}}
  \simeq\frac{kT_2}{m_e c^2}\simeq 5\times10^{-3},
\end{equation}
where $kT_2=2.4\,\rmn{keV}$.  We can now compare the ratio of the Alfv{\'e}n
velocity, $v_{\rmn{A}}$, in the surface layer of the torus to the velocity in
the vortex flow around the minor circle of the torus, $v_{2,\parallel}$,
\begin{equation}
  \label{eq:v_A}
  \frac{v_{\rmn{A}}}{v_{2,\parallel}}
  =\frac{B_\rmn{eq}}{\sqrt{4 \pi \delta \rho_\rmn{2,ICM}^{} v_{2,\parallel}^2}} 
  \simeq 4-11,
\end{equation}
where $n_\rmn{2,ICM} = 2\times 10^{-3}\,\rmn{cm}^{-3}$. In such a configuration,
magnetic tension is strong enough to suppress the Kelvin-Helmholtz instability
of the vortex flow \citep{1961hhs..book.....C, 2007ApJ...670..221D}. To
summarize, if the magnetic field is not too far from equipartition values,
$B\gtrsim B_\rmn{2,eq}/10$, then the line of arguments presented here guarantees
stability of the torus for at least an eddy turnover timescale
$\tau_\rmn{eddy}$.

\section{Discussion and Conclusions}
\label{sec:conclusions}

In providing a quantitative 3D model for NGC 1265, we provide conclusive
observational evidence for an accretion shock onto a galaxy cluster.  The
accretion shock is characterized by an intermediate-strength shock with a Mach
number of $\M\simeq 4.2_{-1.2}^{+0.8}$. In the context of the IGM, it appears
that even weak to intermediate strength shocks of $\M \sim 2.3$ are able to
accelerate electrons as the radio relic in A521 suggests
\citep{2008A&A...486..347G}.  This supports the view that these intermediate
strength shocks are very important in understanding non-thermal processes in
clusters; in particular as they are much more abundant than high-Mach number
shocks \citep[e.g.,][]{2003ApJ...593..599R, 2006MNRAS.367..113P}. The presented
technique offers a novel way to find large-scale formation shock waves that
would likely be missed if we only saw giant radio relics. These are thought to
be due to shock-accelerated relativistic electrons at merging or accretion shock
waves \citep{1998A&A...332..395E, 2006Sci...314..791B}, whereas the
discrimination of the two types of shock waves is difficult on theoretical as
well as observational grounds \citep{2006Sci...314.772E}.

The rich morphology of the large scale radio structure that arches around
steep-spectrum tail of NGC 1265 enables a qualitative argument for the need of
shear to explain the orientation of its ellipsoidally shaped torus.  The energy
density of the shear flow corresponds to a turbulent-to-thermal energy density
of 14\%---consistent with estimates in cosmological simulations.  However,
simulations are needed to quantify the effect of magnetic helicity internal to
the bubble on the torus morphology. In any case, the presence of post-shock
shear implies the amplification of weaker seed magnetic fields through shearing
motions as well as the generation of magnetic fields through the Biermann
battery mechanism if the vorticity has been generated at the curved shock. Even
more interesting, the shock passage of multiphase gas with a large density
contrast necessarily yields a multiply curved shock surface that causes
vorticity injection into later accreted gas on the corresponding curvature
scales. Hence, shearing motions are superposed on various scales which implies
amplification of the magnetic fields on these scales.

More sensitive polarized and total intensity observations of NGC 1265 are needed
to confirm our model predictions with radial polarization vectors of the radio
torus and a polarization degree of around 5\%. In addition, future observations
of different systems similar to NGC 1265 would be beneficial to get a
statistical measurement of properties of the accretions shocks and its pre-shock
conditions. This provides indirect evidence for the existence of the warm-hot
IGM.  Our work shows the potential of these kind of
serendipitous events as ways to explore the outer fringes of clusters and
dynamical features of accretion shocks that are complementary to X-ray
observations.


\acknowledgments We thank T.~A.~En{\ss}lin, T.~Pfrommer, M.~Sun, and an
anonymous referee for helpful comments on this manuscript and gratefully
acknowledge the great atmosphere at the Kavli Institute for Theoretical Physics
program on Particle Acceleration in Astrophysical Plasmas, in Santa Barbara
(2009 July 26--October 3) where this project was initiated.  That program was
supported in part by the National Science Foundation under grant no.
PHY05-51164.  C.P. gratefully acknowledges financial support of the Klaus
Tschira Foundation and the National Science and Engineering Research Council of
Canada. T.W.J. was supported in part by NSF grant AST0908668 and by the University
of Minnesota Supercomputing Institute.


\appendix

\section{Derivation of the jet curvature}

Although fairly standard, for completeness we show the derivation of the jet
curvature radius that is caused by an external ram pressure wind due to the
motion of the galaxy through the ICM. We denote the mass density, velocity, and
radius of the jet by $\rho_\rmn{jet}$, $v_\rmn{jet}$, and $r_\rmn{jet}$,
respectively.  The two jets coming out of the active galactic core back to back
are assumed to be initially a cylinder of length $l_\rmn{jet}$. Each jet is then
bent over a bending radius $r_b $ by the ram pressure wind of mass density and
velocity $\rho_\rmn{ICM}$ and $v$. We equate the jet momentum $\rho_\rmn{jet}
v_\rmn{jet} \pi r_\rmn{jet}^2 l_\rmn{jet}$ with the transverse force due to the
ram pressure wind that acts over a jet propagation timescale (along the bended
path in steady state), $\rho_\rmn{ICM}v^2 2 r_\rmn{jet} l_\rmn{jet} \pi r_b
/(2v_\rmn{jet})$, to obtain the following equality
\begin{equation}
  \label{eq:bending1}
  \rho_\rmn{ICM} v^2\,\frac{\pi r_b }{2 v_\rmn{jet}} = 
  \rho_\rmn{jet} v_\rmn{jet} r_\rmn{jet}\,\frac{\pi}{2}.
\end{equation}
Solving for the ratio of bending-to-jet radius, we obtain
\begin{equation}
  \label{eq:bending2}
  \frac{r_b }{r_\rmn{jet}} = 
  \frac{\M_\rmn{jet}^2}{\M_\rmn{gal}^2}\,
  \frac{\gamma_\rmn{jet}P_\rmn{jet}}{\gamma_\rmn{ICM}P_\rmn{ICM}},
\end{equation}
where we introduced the Mach numbers of the jet and the galaxy,
$\M_\rmn{jet}=v_\rmn{jet}/c_\rmn{jet}$ and $\M_\rmn{gal}=v/c_\rmn{ICM}$, the
adiabatic exponents of jet and surrounding ICM, $\gamma_\rmn{jet}=4/3$ and
$\gamma_\rmn{ICM}=5/3$, and their pressures, $P_\rmn{jet}$ and $P_\rmn{ICM}$.

\section{Derivation of the transformation time of a shocked bubble}
\label{sec:Riemann}

When a low-density bubble crosses a shock of speed $v_\rmn{si}$ in the ICM, the
shock accelerates into the bubble, pulling post-shock ambient gas with it. The
original bubble-ICM CD follows the shock at a speed
intermediate to that of the incident and bubble shocks when the incident shock
is at least moderately strong.  Because the shock intrusion begins sooner and
impacts more strongly on the leading edge of a round bubble than its periphery,
the ambient gas penetrates the center of the bubble first.  An initially
spheroidal bubble will then evolve into a torus (vortex ring) on a timescale
determined by the crossing time of the CD through the bubble.

The shock-produced dynamics in the bubble can be estimated from the exact
solution to the simple 1D Riemann problem of a plane shock impacting a CD.  This
is found most simply in the initial rest frame of the bubble, which we assume to
be in pressure equilibrium with the unshocked ICM and has a density, $\rho_b =
\delta \rho_i$, with $\delta <1$.  At impact a forward shock will penetrate into
the low density bubble with speed, $v_\rmn{sb}$, while a rarefaction will propagate
backward into the post-shock ICM. The bubble-ICM CD will move forward at the
same speed as the post-shock flow in the bubble.  The full Riemann solution is
obtained by matching the pressure behind the forward shock inside the bubble to
the pressure at the foot of the rarefaction in the ICM.

We define the ICM and bubble adiabatic indices as $\gamma_i$ and $\gamma_b$
respectively.  Assuming the shock is propagating from the left, we have right to
left four uniform states, 0, 1, 2, 3, where 0 and 1 are separated by the forward
shock, 1 and 2 are separated by the CD, while 2 and 3 are separated by the
reverse rarefaction (note that this numbering scheme differs from the main body
of the paper). The state ``0'' represents the initial conditions in the bubble,
while ``3'' represents conditions in the ICM post-shock flow.  We set $P_1 = P_2 =
P_*$ and $v_1 = v_2 = v_*=v_\rmn{CD}$.  The initial bubble (and ICM) pressure is
$P_0$, while the initial bubble and ICM sound speeds are related by $c_b = c_0 =
\sqrt{\gamma_b P_0/\rho_0} = c_i\sqrt{\gamma_b/ (\gamma_i \delta)}$.  The Mach
numbers of the external, ICM shock and the internal, bubble shock are
$\mathcal{M}_i = v_\rmn{si}/c_i$ and $\mathcal{M}_b = v_\rmn{sb}/c_b \equiv \mu
\mathcal{M}_i$ and we introduced the Mach number ratio $\mu$.\footnote{Note
  that the variable $\mu$ has a different meaning compared to the main part of
  the paper where it denotes the mean molecular weight.}
\begin{figure*}
\begin{minipage}{\columnwidth}
\begin{center}
  \includegraphics[width=0.45\columnwidth]{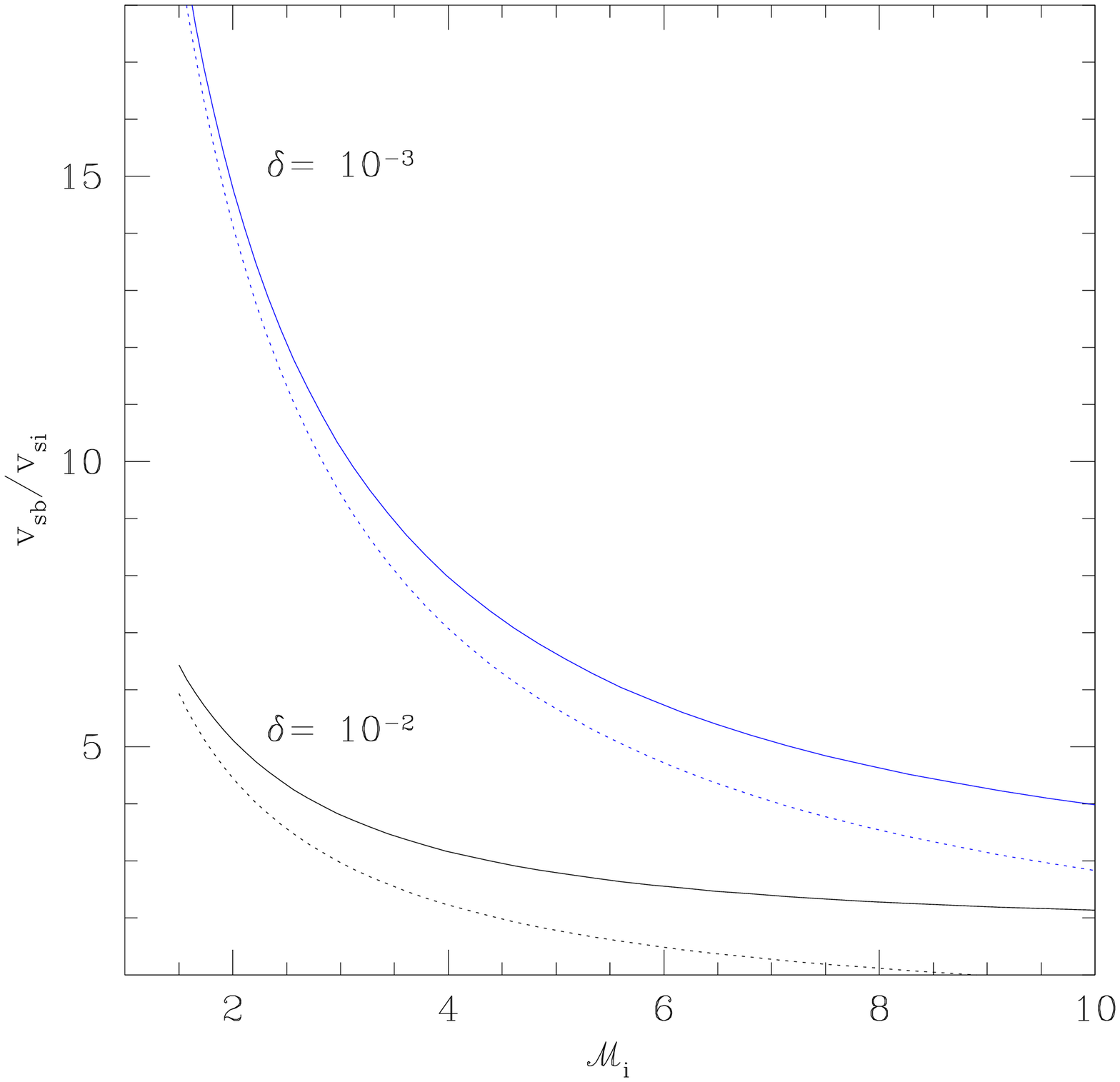}
  \includegraphics[width=0.45\columnwidth]{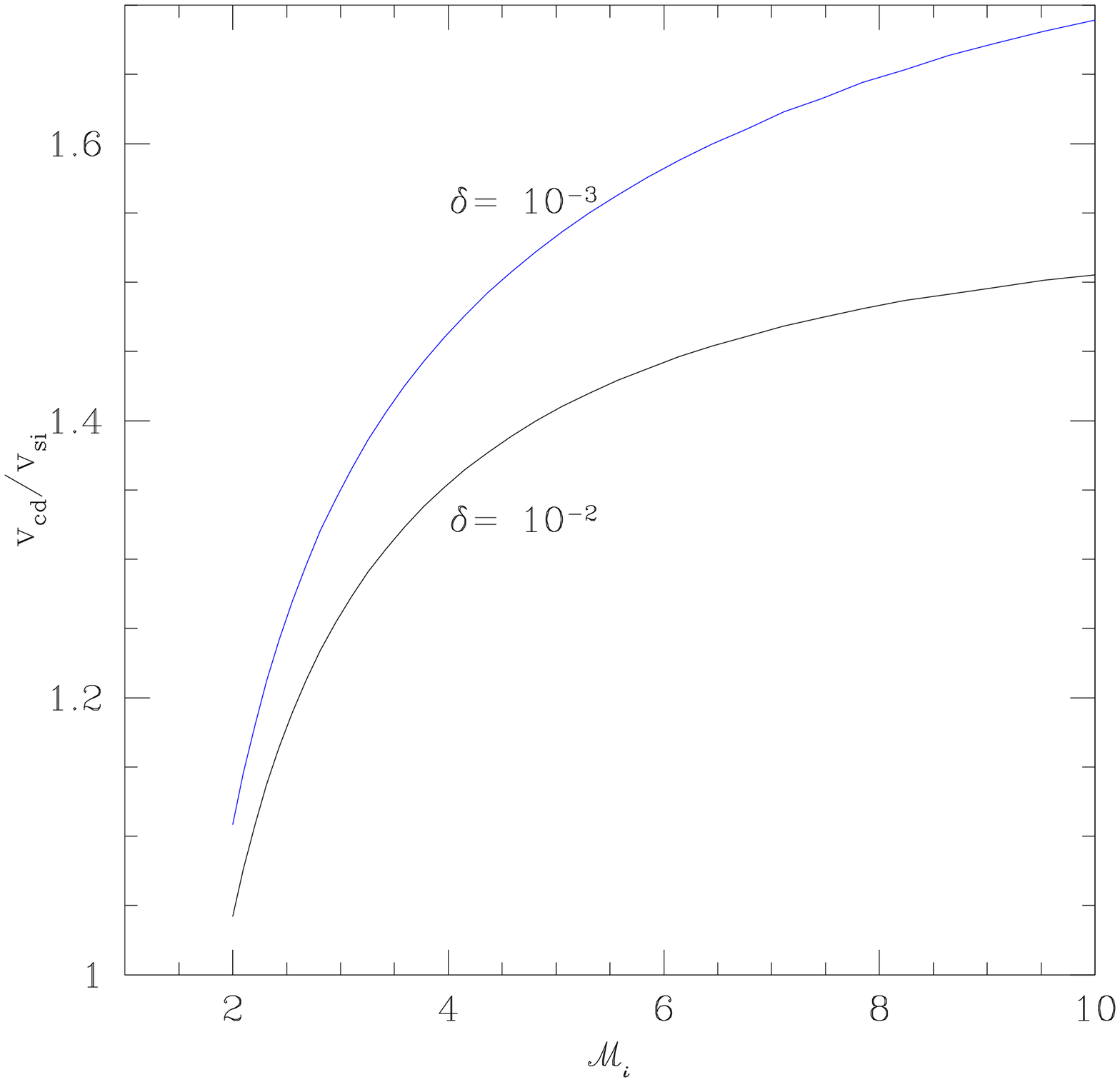}
\end{center}
\caption{Left: ratio of the speed of a shock inside a low density bubble,
  $v_\rmn{sb}$, to the incident shock speed, $v_\rmn{si}$, for density contrasts,
  $\delta = 10^{-2},~10^{-3}$.  Solid curves represent solutions to
  Equation~(\ref{eq:B5}). Dotted curves represent the internal bubble sound speed
  in units of the external shock speed in the ICM.  Right: ratio of the contact
  discontinuity speed inside the bubble to the incident shock speed from the
  relevant expression in Equation~(\ref{eq:B1}).}
\label{bubshock}
\end{minipage}
\end{figure*}

From standard shock jump conditions, we have
\begin{eqnarray}
P_3 = \frac{2}{\gamma_i+1}\left(\gamma_i \mathcal{M}_i^2 - \frac{\gamma_i - 1}{2}\right)P_0\nonumber\\
\rho_3 = \frac{(\gamma_i+1)\mathcal{M}_i^2}{(\gamma_i-1)\mathcal{M}_i^2 + 2}\rho_i 
=\frac{1}{\delta}\frac{(\gamma_i+1)\mathcal{M}_i^2}{(\gamma_i-1)\mathcal{M}_i^2+2}\rho_b\nonumber\\
v_3 = \frac{2}{\gamma_i+1}\frac{(\mathcal{M}_i^2 - 1)}{\mathcal{M}_i}c_{i}\\
P_1 = P_2 = P_* = \frac{2}{\gamma_b+1}\left(\gamma_b\mu^2\mathcal{M}_i^2 - \frac{\gamma_b-1}{2}\right)P_0\nonumber\\
\rho_1=\frac{(\gamma_b+1)\mu^2\mathcal{M}_i^2}{(\gamma_b-1)\mu^2\mathcal{M}_i^2 + 2}\rho_b\nonumber\\
v_\rmn{CD} = v_1 = v_2 = v_* = \frac{2}{\gamma_b+1}\frac{(\mu^2\mathcal{M}_i^2 - 1)}{\mu\mathcal{M}_i}c_b\nonumber.
\label{eq:B1}
\end{eqnarray}
In addition, the Riemann invariant connecting states 2 and 3 through the
rarefaction, along with the relation $\rho \propto P^{1/\gamma_i}$ inside
the rarefaction give
\begin{equation}
v_* = v_3 + \frac{2}{\gamma_i-1} c_3 \left[1 - \left(\frac{P_*}{P_3}\right)^{\frac{\gamma_i-1}{2\gamma_i}}\right],
\end{equation}
while 
\begin{equation}
c_3^2 = \frac{\gamma_i P_3}{\rho_3} = 
\frac{2}{\gamma_i+1}\frac{\left[\gamma_i\mathcal{M}_i^2- \frac{\gamma_i-1}{2}\right]\left[(\gamma_i-1)\mathcal{M}_i^2 + 2\right]}{(\gamma_i+1)\mathcal{M}_i^2}c_i^2.
\end{equation}
These can be combined to give the following equation for $\mu = \mathcal{M}_b/\mathcal{M}_i$: 
\begin{eqnarray}
1=\frac{\gamma_i+1}{\gamma_b+1}\frac{\mu^2\mathcal{M}_i^2-1}{\mu\delta^{1/2}(\mathcal{M}_i^2-1)}\left(\frac{\gamma_b}{\gamma_i}\right)^{1/2}\nonumber\\
-\frac{1}{\gamma_i-1}\frac{\left[2\gamma_i\mathcal{M}_i^2-\gamma_i+1\right]^{1/2}\left[(\gamma_i-1)\mathcal{M}_i^2+2\right]^{1/2}}{\mathcal{M}_i^2-1}
\left[1-\left(\frac{(\gamma_i+1)(2\gamma_b\mu^2\mathcal{M}_i^2-\gamma_b+1)}{(\gamma_b+1)(2\gamma_i\mathcal{M}_i^2-\gamma_i+1)}\right)^{\frac{\gamma_i-1}{2\gamma_i}}\right]
\end{eqnarray}
which can be solved numerically for $\mu$, given $\mathcal{M}_i$, $\gamma_i$, $\gamma_b$, and $\delta$.
For $\delta = 1$, $\gamma_i=\gamma_b$ the obvious solution is $\mu = 1$. 
For our problem we expect $\gamma_i = 5/3$ and $\gamma_b = 4/3$. This gives
\begin{equation}
\label{eq:B5}
1=\frac{8}{7}\left(\frac{4}{5}\right)^{1/2}\frac{\mu^2\mathcal{M}_i^2-1}{\mu\delta^{1/2}(\mathcal{M}_i^2-1)}
-\frac{\left[(5\mathcal{M}_i^2-1)(\mathcal{M}_i^2+3)\right]^{1/2}}{\mathcal{M}_i^2-1}
\left[1-\left(\frac{4}{7}\right)^{1/5}\left(\frac{8\mu^2\mathcal{M}_i^2-1}{5\mathcal{M}_i^2-1}\right)^{1/5}\right].
\end{equation}
Equation~(\ref{eq:B5}) is displayed for $\delta = 10^{-2}$ and $\delta = 10^{-3}$ in
the left panel of Figure~\ref{bubshock}.


In the strong (external) shock limit, $\mu\mathcal{M}_i \gg 1$, it is easy to
see that approximately $\mu \propto \sqrt{\delta}$. Empirically we find in this
limit when $\delta \ll 1$ that $\mu \sim 2 \sqrt{\delta}$, so that $v_\rmn{sb}
\approx 2 v_\rmn{si}$. As $\mathcal{M}_i \rightarrow 1$, the solution $\mu
\rightarrow 1$, so $\mathcal{M}_b \rightarrow 1$ also applies. The more general
solutions for small to moderate incident shock strengths are shown for two
values of $\delta$ as solid curves in Figure~\ref{bubshock}. The lower bound for
$v_\rmn{sb}$ is given by $c_b =
v_\rmn{si}\sqrt{\gamma_b/\gamma_i}~[1/(\mathcal{M}_i\sqrt{\delta})]$ and is shown in
each case by a dotted curve.  We note that the solution shown in
Figure~\ref{bubshock} is almost identical to the case of $\gamma_i =\gamma_b$,
since the internal shock in the bubble is barely supersonic with $\mathcal{M}_b
\gtrsim 1$ due to the large sound speed in the bubble. This can be easily seen
by taking the ratio of the corresponding solid-to-dotted lines which provides
$\mathcal{M}_b$ for each external Mach number $\mathcal{M}_i$.

We can understand the general behavior of an increasing
shock speed inside a low-density bubble relative to the incident shock speed for
smaller Mach numbers $\mathcal{M}_i$ by the following line of arguments. For
small Mach numbers, both waves are just nonlinear sound waves, so each
propagates near the local sound speed, which is much larger in the bubble. It
turns out for large Mach numbers $\mathcal{M}_i$ that the pressure just behind
the penetrating shock is smaller than the external post-shock pressure by a
factor that scales with the density contrast $\delta$.  This pressure drop comes
from the rarefaction going back into the shocked ICM. For weak shocks that
rarefaction is weak (provided the original bubble was in pressure equilibrium).

The right panel in Figure~\ref{bubshock} shows the behavior of $v_\rmn{CD}$
corresponding to the shock solutions in the left figure panel. In the limit
$\mu^2 \mathcal{M}_i^2\gg 1$, the expression for $v_\rmn{CD}$ in Equation~(\ref{eq:B1}) takes
the form $v_\rmn{CD}/v_\rmn{si} \approx (2/(\gamma_b+1))
\sqrt{\gamma_b/\delta\gamma_i}~\mu$.  Note that the limit $\mu^2
\mathcal{M}_i^2\gg 1$ requires the internal bubble (pseudo-) temperature not to
be too high and hence implies a lower limit on the bubble density contrast of
$\delta\gtrsim 10^{-3}$ assuming the original bubble was in pressure
equilibrium.  Applying our empirical result for strong shocks with
$10^{-3}\lesssim\delta \ll 1$ (namely, $\mu \sim 2 \sqrt{\delta}$), we would
expect $v_\rmn{CD}/v_\rmn{si}\rightarrow (4/(\gamma_b+1))\sqrt{\gamma_b/\gamma_i} \sim
1.5$, which is within about 20\% of the exact solutions in this limit. This
value for $v_\rmn{CD}/v_\rmn{si}$ is also a reasonable estimate at moderate shock
numbers, say $\mathcal{M}_i\ga 3$.  For $\mathcal{M}_i\ga 2$ the limit $v_\rmn{CD}>
v_\rmn{si}$ still applies.  As $\mathcal{M}_i$ approaches one, however, the speed of
the CD drops below the incident shock speed, since as the bubble shock becomes a
sound wave, the CD's motion vanishes. The results of these considerations are
quantitatively confirmed by a suite of 2D axisymmetric simulations where we
varied the Mach number $\mathcal{M}_i = \{2,3,5,10\}$ and initial bubble/ICM
density contrast (see Figure~\ref{fig:images} for one realization).

For our moderate-strength shock example in Perseus we can use these results to
estimate the time for the CD to cross a bubble diameter, $2R_\rmn{bubble}$,
transforming the bubble into a vortex ring, with a toroidal topology; namely,
\begin{equation}
\tau_{\rm{form}}\sim \frac{2 R_\rmn{bubble}}{1.5 v_\rmn{si}}\sim 1.4\times 10^{8}~\rmn{yr}.
\label{eq:tau_form}
\end{equation}
Here we adopted our fiducial values of $R_\rmn{bubble}\simeq150\,\rmn{kpc}$,
$\delta\simeq 5\times10^{-3}$ (Equation~(\ref{eq:delta})), and $v_\rmn{si} =
\mathcal{M}_i c_i = \mathcal{M}_i \sqrt{\gamma kT_i / (\mu_\rmn{mw} m_p)} \simeq
4.2\times 330\,\rmn{km~s}^{-1} \simeq 1400\,\rmn{km~s}^{-1}$ where we used the sound speed
in the pre-shock gas in the infalling filament of temperature
$kT_i\simeq0.4\,\rmn{keV}$ and $\mu_\rmn{mw}=0.588$. We emphasize that this estimate
is close to the value $\tau_{3\to4} = 1.2\times 10^8\,\rmn{yr}$ that we adopted in
our model and consistent with the derived the error bars on $\mathcal{M}_i$.

\bibliographystyle{apj}
\bibliography{bibtex/paper}

\end{document}